\documentclass[a4paper,10pt]{revtex4}
\usepackage{graphicx}  
\usepackage{amsmath}   
\usepackage{amssymb}   
\usepackage{bm} 
\usepackage{dcolumn}
\usepackage{color}
\usepackage{mathrsfs}
\usepackage{amsfonts}
\usepackage{varioref}
\usepackage{mathrsfs}
\usepackage{graphicx}
\usepackage{latexsym}
\usepackage{amsmath}
\usepackage{amssymb}
\usepackage{textcomp}
\usepackage{amsbsy}
\usepackage{graphics}
\usepackage{epstopdf}
\usepackage{color}
\usepackage[caption=false]{subfig}

\RequirePackage[colorlinks,citecolor=blue,urlcolor=magenta,linkcolor=blue]{hyperref}
\input epsf

\allowdisplaybreaks[4]

\begin{document}

\tolerance=5000

\title{Early and late universe holographic cosmology from a new generalized entropy}

\author{Shin'ichi~Nojiri$^{1,2}$\,\thanks{nojiri@gravity.phys.nagoya-u.ac.jp},
Sergei~D.~Odintsov$^{3,4}$\,\thanks{odintsov@ieec.uab.es},
Tanmoy~Paul$^{5}$\,\thanks{pul.tnmy9@gmail.com}} \affiliation{
$^{1)}$ Department of Physics, Nagoya University,
Nagoya 464-8602, Japan \\
$^{2)}$ Kobayashi-Maskawa Institute for the Origin of Particles
and the Universe, Nagoya University, Nagoya 464-8602, Japan \\
$^{3)}$ ICREA, Passeig Luis Companys, 23, 08010 Barcelona, Spain\\
$^{4)}$ Institute of Space Sciences (ICE, CSIC) C. Can Magrans s/n, 08193 Barcelona, Spain\\
$^{5)}$ Department of Physics, Chandernagore College, Hooghly - 712 136, India.}


\tolerance=5000

\begin{abstract}
We propose a new four-parameter entropy function that generalizes the Tsallis, R\'{e}nyi, Barrow, Sharma-Mittal, 
Kaniadakis and Loop Quantum Gravity entropies for suitable limits of the parameters. 
Consequently, we address the early and late universe cosmology corresponding to the proposed four-parameter entropy function. 
As a result, it turns out that the entropic cosmology from the generalized entropy function can 
unify the early inflation to the late dark energy era of the universe. 
In such a unified scenario, we find that -- 
(1) the inflation era is described by a quasi de-Sitter evolution of the Hubble parameter, which has an exit at 
around 58 e-folding number, (2) the inflationary observable quantities like the spectral index 
for primordial scalar perturbation and the tensor-to-scalar ratio are simultaneously compatible with the recent Planck data, and 
(3) regarding the late time cosmology, the dark energy EoS parameter is found to be consistent with the Planck result for 
the same values of the entropy parameters that lead to the viable inflation during the early universe. 
Furthermore, we show that the entropic cosmology from the proposed 
entropy function is equivalent to holographic cosmology, where the respective holographic cut-offs are determined in terms of either particle horizon 
and its derivative or future horizon and its derivative. 
\end{abstract}

\maketitle

\section{Introduction}

Entropy is regarded as one of the fundamental quantities and plays a significant role in physics since the modern days. 
However with the advancement of quantum field theory, and quantum gravity, it seems that that entropy may not be unique, 
rather it depends on the physical system under consideration. 
Or possibly it indicates that we do not fundamentally understand what the physical entropy is; maybe, there exists a generalized 
form of entropy that can be applied irrespective of physical systems.

One of the important discoveries in theoretical physics is the black body radiation of a black hole, which is described by a 
certain temperature and a Bekenstein-Hawking entropy \cite{Bekenstein:1973ur,Hawking:1975vcx} (see \cite{Bardeen:1973gs,Wald:1999vt} for reviews). 
However unlike classical thermodynamics where the entropy is directly proportional 
to the volume of the system, the Bekenstein-Hawking entropy is proportional to the horizon area. 
Based on such distinctive nature of entropy, recent literatures proposed various forms of entropy other than the Bekenstein-Hawking one, 
depending on the non-additive statistics, like the Tsallis \cite{Tsallis:1987eu} and the R\'{e}nyi \cite{Renyi} entropies. 
Recently the Barrow entropy was proposed in \cite{Barrow:2020tzx}, which may encode the fractal features of a black hole structure that originated 
from quantum gravity effects. 
Some other well-known entropies are the Sharma-Mittal entropy \cite{SayahianJahromi:2018irq} 
(which can be seen as a possible combination of the Tsallis and the R\'{e}nyi entropies), the Kaniadakis entropy \cite{Kaniadakis:2005zk,Drepanou:2021jiv} 
and the entropy in the context of Loop Quantum Gravity \cite{Majhi:2017zao,Liu:2021dvj} etc. 
All of such entropies share some common properties, like -- they reduce to the Bekenstein-Hawking entropy at a 
certain limit and they are monotonically increasing function with respect to the Bekenstein-Hawking entropy variable. 
This may raise the possibility that there exists a more generalized entropy function that can generalize all the known aforementioned entropies. 

Furthermore the horizon thermodynamics is extended to the cosmology sector. 
In this regard holographic cosmology, initiated by Witten and Susskind in \cite{Witten:1998qj,Susskind:1998dq,Fischler:1998st}, 
earned a lot of attention as it is directly related to the entropy construction. 
Recently, we showed that the entropic cosmology corresponding to various entropic functions are 
equivalent to holographic cosmology with suitable holographic cut-offs \cite{Nojiri:2021iko,Nojiri:2021jxf}. 
The most intriguing question in modern cosmology is to explain the accelerating phases of the universe during 
two extreme curvature regimes, namely the inflation and the dark energy era of the universe. 
The holographic cosmology sourced from the aforementioned entropies successfully explain the dark energy era of the universe for constant as well as 
for variable exponents of the entropy functions, and generally known as the holographic dark energy (HDE) model 
\cite{Li:2004rb,Li:2011sd,Wang:2016och,Pavon:2005yx,Nojiri:2005pu,Enqvist:2004xv,Zhang:2005yz,Guberina:2005fb,Elizalde:2005ju,
Ito:2004qi,Gong:2004cb,Saridakis:2007cy,Gong:2009dc,BouhmadiLopez:2011xi,Malekjani:2012bw,Khurshudyan:2014axa,Khurshudyan:2016gmb,Landim:2015hqa,
Gao:2007ep,Li:2008zq,Anagnostopoulos:2020ctz,Zhang:2005hs,Li:2009bn,Feng:2007wn,Zhang:2009un,Lu:2009iv,
Micheletti:2009jy,Huang:2004wt,Mukherjee:2017oom,Nojiri:2017opc,Nojiri:2019skr,
Saridakis:2020zol,Barrow:2020kug,Adhikary:2021xym,Srivastava:2020cyk,Bhardwaj:2021chg,Chakraborty:2020jsq,Sarkar:2021izd}. 
Besides the dark energy era, the holographic cosmology is also able to trigger a viable inflationary scenario during the early stage of the universe 
\cite{Nojiri:2022aof,Horvat:2011wr,Nojiri:2019kkp,Paul:2019hys,Bargach:2019pst,Elizalde:2019jmh,Oliveros:2019rnq,Mohammadi:2022vru,Chakraborty:2020tge}, 
and more interestingly, the holographic cosmology can unify an early inflation to the late dark energy era in a covariant formalism \cite{Nojiri:2020wmh}. 
Actually, during the early phase when the size of the universe is small, the holographic energy density is huge and thus triggers an inflationary scenario 
that is consistent with the recent Planck data. 
From a different viewpoint, the holographic scenario is extended to explain a bouncing cosmology, in which case, the holographic energy density 
helps to violate the null energy condition around the bounce \cite{Nojiri:2019yzg,Brevik:2019mah}. 
All of these works indicate the intense interest on holographic or equivalently on 
entropic cosmology corresponding to various entropy functions. 

Based on the arguments in \cite{Nojiri:2022aof}, the following questions are raised:
\begin{itemize}
\item Does there exist a generalized form of entropy that generalizes the known entropy functions like the 
Tsallis, R\'{e}nyi, Barrow, Sharma-Mittal, Kaniadakis and Loop Quantum Gravity entropies?
\item If so, then what are the cosmological implications of such generalized entropy function?
\end{itemize}
The answer to these questions is affirmative, particularly in \cite{Nojiri:2022aof}, a 6 parameter generalized entropy construction has been proposed. 
However, unlike to \cite{Nojiri:2022aof}, we will propose 
a different entropy function containing fewer parameters, particularly having 4 parameters, 
which can generalize all the known entropies mentioned. 
In this regard, we will give the following postulate: ``The minimum number of parameters required in a generalized entropy function that can generalize 
all the aforementioned entropies is equal to four''. Consequently, we will address the early and late time cosmological implications 
of the universe from the proposed four-parameter entropy function.

The paper is organized as follows: in Sec.~\ref{sec-gen-entropy}, we will provide the generalized four parameter entropy function. 
The modified Friedmann equations corresponding to such entropy will be shown in Sec.~\ref{SecI}. 
The holographic equivalence of the proposed entropy will be established 
in Sec.~\ref{sec-equivalence}. The cosmology from the generalized entropy function will be presented in Sec.~\ref{sec-inf} 
and Sec.~\ref{sec-DE}. 
Finally, the paper ends with some conclusions.

\section{A generalized four parameter entropy}\label{sec-gen-entropy}

Here we will propose a generalized four-parameter entropy function which can lead to various known entropy functions proposed so far 
for suitable choices of the parameters.

Let us start with the Bekenstein-Hawking entropy, the very first proposal of thermodynamical entropy of 
black hole physics \cite{Bekenstein:1973ur,Hawking:1975vcx},
\begin{align}
S = \frac{A}{4G} \,,
\label{BH-entropy}
\end{align}
where $A = 4\pi r_h^2$ is the area of the horizon and $r_h$ is the horizon radius. 
Consequently, different entropy functions have been introduced 
depending on the system under consideration. Let us briefly recall some of the entropy functions proposed so far:

\begin{itemize}
\item For the systems with long range interactions where the Boltzmann-Gibbs entropy is not applied, one needs to introduce a generalized entropy known as 
the Tsallis entropy which is given by \cite{Tsallis:1987eu},
\begin{align}
S_T = \frac{A_0}{4G}\left(\frac{A}{A_0}\right)^{\delta} \,,
\label{Tsallis entropy}
\end{align}
where $A_0$ is a constant and $\delta$ is the exponent. 
It is evident that the Tsallis entropy converges to the Bekenstein-Hawking entropy for $\delta = 1$. 
\item The R\'{e}nyi entropy is given by \cite{Renyi},
\begin{align}
S_\mathrm{R} = \frac{1}{\alpha} \ln \left( 1 + \alpha S \right) \,,
\label{Renyi entropy}
\end{align}
where $S$ is identified with the Bekenstein-Hawking entropy and $\alpha$ is a parameter. 
Clearly for $\alpha \rightarrow 0$, the R\'{e}nyi entropy function resembles with the Bekenstein-Hawking one. 
\item The Barrow entropy is given by \cite{Barrow:2020tzx},
\begin{align}
\label{Barrow-entropy}
S_\mathrm{B} = \left(\frac{A}{A_\mathrm{Pl}} \right)^{1+\Delta/2} \,,
\end{align}
where $A$ is the usual black hole horizon area and $A_\mathrm{Pl} = 4G$ is the Planck area. 
Inspired from the Covid-19 virus illustrations, Barrow argued 
that quantum-gravitational effects may introduce intricate, fractal features on the black-hole structure. 
The case $\Delta = 1$ refers to the most fractal structure of black hole.  
\item The Sharma-Mittal entropy is given by \cite{SayahianJahromi:2018irq},
\begin{align}
S_{SM} = \frac{1}{R}\left[\left(1 + \delta ~S\right)^{R/\delta} - 1\right] \,,
\label{SM entropy}
\end{align}
where $R$ and $\delta$ are two parameters. 
The Sharma-Mittal entropy can be regarded as a possible combination of the Tsallis and R\'{e}nyi entropies. 
\item The Kaniadakis entropy function comes with the following form \cite{Kaniadakis:2005zk,Drepanou:2021jiv}:
\begin{align}
S_K = \frac{1}{K}\sinh{\left(KS\right)} \,,
\label{K-entropy}
\end{align}
where $K$ is a phenomenological parameter. 
The Kaniadakis entropy has a limit of Bekenstein-Hawking entropy at $K \rightarrow 0$. 
\item In the context of Loop Quantum Gravity, one may get the following entropy function 
\cite{Majhi:2017zao,Liu:2021dvj}:
\begin{align}
S_q = \frac{1}{\left(1-q\right)}\left[\mathrm{e}^{(1-q)\Lambda(\gamma_0)S} - 1\right] \,,\label{LQG entropy}
\end{align}
where $q$ is the exponent and $\Lambda(\gamma_0) = \ln{2}/\left(\sqrt{3}\pi\gamma_0\right)$ with $\gamma_0$ being the Barbero-Immirzi parameter. 
The $\gamma_0$ generally takes either $\gamma_0 = \frac{\ln{2}}{\pi\sqrt{3}}$ or $\gamma_0 = \frac{\ln{3}}{2\pi\sqrt{2}}$. 
However with $\gamma_0 = \frac{\ln{2}}{\pi\sqrt{3}}$, $\Lambda(\gamma_0)$ becomes unity 
and $S_q$ resembles with the Bekenstein-Hawking entropy for $q \rightarrow 1$. 
\end{itemize}

\subsection*{Generalized entropy}

At this stage, we propose a new 4 parameter entropy function which can generalize all the entropies mentioned above. In particular, 
the generalized entropy ($S_\mathrm{g}$) is given by,
\begin{align}
S_\mathrm{g}\left[\alpha_+,\alpha_-,\beta,\gamma \right] = \frac{1}{\gamma}\left[\left(1 + \frac{\alpha_+}{\beta}~S\right)^{\beta} 
 - \left(1 + \frac{\alpha_-}{\beta}~S\right)^{-\beta}\right] \,,
\label{gen-entropy}
\end{align}
where $\alpha_+$, $\alpha_-$, $\beta$, and $\gamma$ are the parameters which are assumed to be positive, and $S$ is the Bekenstein-Hawking entropy. 
First we show that the entropy $S_\mathrm{g}\left[ \alpha_+,\alpha_-,\beta,\gamma \right]$ 
reduces to the entropies mentioned in Eqs.~(\ref{Tsallis entropy}), 
(\ref{Renyi entropy}), (\ref{Barrow-entropy}), (\ref{SM entropy}), (\ref{K-entropy}), and (\ref{LQG entropy}) for suitable choices of the parameters. 
\begin{itemize}
\item For $\alpha_+ \rightarrow \infty$ and $\alpha_- = 0$, one gets
\begin{align}
S_\mathrm{g} = \frac{1}{\gamma}\left(\frac{\alpha_+}{\beta}\right)^{\beta}S^{\beta} \,.\nonumber
\end{align}
If we further choose $\gamma = \left(\alpha_+/\beta\right)^{\beta}$, then the generalized entropy reduces to
\begin{align}
 S_\mathrm{g} = S^{\beta} \,.\nonumber
\end{align}
Therefore with $\beta = \delta$ or $\beta = 1 + \Delta$, the generalized entropy resembles with the Tsallis entropy or with the Barrow entropy 
respectively.

\item For $\alpha_- = 0$, $\beta \rightarrow 0$ and $\frac{\alpha_+}{\beta} \rightarrow \mathrm{finite}$ -- Eq.~(\ref{gen-entropy}) leads to,
\begin{align}
 S_\mathrm{g} = \frac{1}{\gamma}\left[\left(1 + \frac{\alpha_+}{\beta}~S\right)^{\beta} - 1\right] 
 = \frac{1}{\gamma}\left[\exp{\left\{\beta\ln{\left(1 + \frac{\alpha_+}{\beta}~S\right)}\right\}} - 1\right] 
 \approx \frac{1}{\left(\gamma/\beta\right)} \ln{\left(1 + \frac{\alpha_+}{\beta}~S\right)} \,.\nonumber
\end{align}
Further choosing $\gamma = \alpha_+$ and identifying $\frac{\alpha_+}{\beta} = \alpha$, we can write the above expression as,
\begin{align}
 S_\mathrm{g} = \frac{1}{\alpha}\ln{\left(1 + \alpha~S\right)} \,,
\end{align}
i.e.,  $S_\mathrm{g}$ reduces to the R\'{e}nyi entropy.

\item In the case when $\alpha_- = 0$, the generalized entropy becomes,
\begin{align}
 S_\mathrm{g} = \frac{1}{\gamma}\left[\left(1 + \frac{\alpha_+}{\beta}~S\right)^{\beta} - 1\right] \,.
\end{align}
Thereby identifying $\gamma = R$, $\alpha_+ = R$ and $\beta = R/\delta$, the generalized entropy function $S_\mathrm{g}$ gets similar to the 
Sharma-Mittal entropy.

\item For $\beta \rightarrow \infty$, $\alpha_+ = \alpha_- = \frac{\gamma}{2} = K$, we may write Eq.~(\ref{gen-entropy}) as,
\begin{align}
S_\mathrm{g}=&\, \frac{1}{2K}\lim_{\beta \rightarrow \infty}\left[\left(1 + \frac{K}{\beta}~S\right)^{\beta} 
 - \left(1 + \frac{K}{\beta}~S\right)^{-\beta}\right]\nonumber\\
=&\, \frac{1}{2K}\left[\mathrm{e}^{KS} - \mathrm{e}^{-KS}\right] = \frac{1}{K}\sinh{\left(KS\right)} 
\rightarrow \mathrm{Kaniadakis~entropy} \,.
\end{align}

\item Finally, with $\alpha_- = 0$, $\beta \rightarrow \infty$ and $\gamma = \alpha_+ = (1-q)$, Eq.~(\ref{gen-entropy}) immediately yields,
\begin{align}
 S_\mathrm{g} = \frac{1}{(1-q)}\left[\mathrm{e}^{(1-q)S} - 1\right] \,,\nonumber
\end{align}
which is the Loop Quantum Gravity entropy with $\Lambda(\gamma_0) = 1$ or equivalently $\gamma_0 = \frac{\ln{2}}{\pi\sqrt{3}}$.

\end{itemize}
Furthermore, the generalized entropy function in Eq.~(\ref{gen-entropy}) have the following properties: 
(1) $S_\mathrm{g}\left[ \alpha_+,\alpha_-,\beta,\gamma \right]$ 
satisfies the generalized third law of thermodynamics, i.e.,  $S_\mathrm{g} \rightarrow 0$ for $S \rightarrow 0$. (2) The entropy 
$S_\mathrm{g}\left[ \alpha_+,\alpha_-,\beta,\gamma \right]$ is a monotonically increasing function with $S$ because both the terms 
$\left(1 + \frac{\alpha_+}{\beta}~S\right)^{\beta}$ and $-\left(1 + \frac{\alpha_-}{\beta}~S\right)^{-\beta}$ present in the expression 
of $S_\mathrm{g}$ 
increase with $S$. (3) $S_\mathrm{g}\left[ \alpha_+,\alpha_-,\beta,\gamma \right]$ seems to converge to the Bekenstein-Hawking 
entropy at certain limit of the parameters. 
In particular, for $\alpha_+ \rightarrow \infty$, $\alpha_- = 0$, $\gamma = \left(\alpha_+/\beta\right)^{\beta}$ and $\beta = 1$, 
the generalized entropy function 
in Eq.~(\ref{gen-entropy}) becomes equivalent to the Bekenstein-Hawking entropy.

Here it deserves mentioning that beside the entropy function proposed in Eq.~(\ref{gen-entropy}) which contains four parameters, one may consider 
a three parameter entropy having the following form:
\begin{align}
S_3[\alpha,\beta,\gamma] = \frac{1}{\gamma}\left[\left(1 + \frac{\alpha}{\beta}~S\right)^{\beta} - 1\right] \,,
\label{3-parameter-entropy}
\end{align}
where $\alpha$, $\beta$ and $\gamma$ are the parameters. 
The above form of $S_3[\alpha,\beta,\gamma]$ satisfies all the properties, 
like -- (1) $S_3[\alpha,\beta,\gamma] \rightarrow 0$ for $S \rightarrow 0$, (2) $S_3$ is an increasing function with $S$ and (3) $S_3$ 
has a Bekenstein-Hawking entropy limit for the choices: $\alpha \rightarrow \infty$, $\gamma = \left(\alpha/\beta\right)^{\beta}$ and $\beta = 1$ 
respectively. 
However $S_3[\alpha,\beta,\gamma]$ is not able to generalize all the known entropies mentioned from Eq.~(\ref{Tsallis entropy}) 
to Eq.~(\ref{LQG entropy}), in particular, $S_3[\alpha,\beta,\gamma]$ does not reduce to the Kaniadakis entropy for any possible 
choices of the parameters.\\

\textbf{\underline{A Postulate}}: 
Based on our findings, we propose the following postulate in regard to the generalized entropy function -- 
``The minimum number of parameters required in a generalized entropy function that can generalize 
all the known entropies mentioned from Eq.~(\ref{Tsallis entropy}) to Eq.~(\ref{LQG entropy}) is equal to four''.

\section{Modified Friedmann equations corresponding to the generalized entropy} \label{SecI}

We will consider the thermodynamic approach to describe the cosmological behaviour of the universe from the four parameter generalized 
entropy function $S_\mathrm{g}$ given in Eq.(\ref{gen-entropy}).

The Friedmann-Lema\^{i}tre-Robertson-Walker space-time with flat spacial part will serve our purpose, in particular,
\begin{align}
ds^2=-dt^2+a^2(t)\sum_{i=1,2,3} \left(dx^i\right)^2 \, .
\label{metric}
\end{align}
Here $a(t)$ is called as a scale factor. 

The radius $r_\mathrm{H}$ of the cosmological horizon is given by 
\begin{align}
\label{apphor}
r_\mathrm{H}=\frac{1}{H}\, ,
\end{align}
with $H = \dot{a}/a$ is the Hubble parameter of the universe. 
Then the entropy contained within the cosmological horizon can be obtained from 
the Bekenstein-Hawking relation \cite{Padmanabhan:2009vy}. 
Furthermore the flux of the energy $E$, or equivalently, the increase of the heat $Q$ in the region comes as 
\begin{align}
\label{Tslls2}
dQ = - dE = -\frac{4\pi}{3} r_\mathrm{H}^3 \dot\rho dt = -\frac{4\pi}{3H^3} \dot\rho~dt 
= \frac{4\pi}{H^2} \left( \rho + p \right)~dt \, ,
\end{align}
where, in the last equality, we use the conservation law: $0 = \dot \rho + 3 H \left( \rho + p \right)$. 
Then from the Hawking temperature \cite{Cai:2005ra}
\begin{align}
\label{Tslls6}
T = \frac{1}{2\pi r_\mathrm{H}} = \frac{H}{2\pi}\, ,
\end{align}
and by using the first law of thermodynamics $TdS = dQ$, 
one obtains $\dot H = - 4\pi G \left( \rho + p \right)$. Integrating the expression immediately leads to 
the first FRW equation, 
\begin{align}
\label{Tslls8}
H^2 = \frac{8\pi G}{3} \rho + \frac{\Lambda}{3} \, ,
\end{align}
where the integration constant $\Lambda$ can be treated as a cosmological constant. 

Instead of the Bekenstein-Hawking entropy of Eq.~(\ref{BH-entropy}), we may 
use the generalized entropy in Eq.~(\ref{gen-entropy}), in regard to which, the first law of thermodynamics leads to the following equation:
\begin{align}
\dot{H}\left(\frac{\partial S_\mathrm{g}}{\partial S}\right) = -4\pi G\left(\rho + p\right) \,.
\label{FRW1-sub}
\end{align}
With the explicit form of $S_\mathrm{g}$ from Eq.~(\ref{gen-entropy}), the above equation turns out to be,
\begin{align}
\frac{1}{\gamma}\left[\alpha_{+}\left(1 + \frac{\pi \alpha_+}{\beta GH^2}\right)^{\beta - 1} 
+ \alpha_-\left(1 + \frac{\pi \alpha_-}{\beta GH^2}\right)^{-\beta-1}\right]\dot{H} = -4\pi G\left(\rho + p\right)
\label{FRW-1}
\end{align}
where we use $S = A/(4G) = \pi/(GH^2)$. 
Using the conservation relation of the matter fields, i.e., $\dot{\rho} + 3H\left(\rho + p\right) = 0$, 
Eq.~(\ref{FRW-1}) can be written as,
\begin{align}
\frac{2}{\gamma}\left[\alpha_{+}\left(1 + \frac{\pi \alpha_+}{\beta GH^2}\right)^{\beta - 1} 
+ \alpha_-\left(1 + \frac{\pi \alpha_-}{\beta GH^2}\right)^{-\beta-1}\right]H~dH = \left(\frac{8\pi G}{3}\right)d\rho \,,
 \nonumber
\end{align}
on integrating which, we obtain, 
\begin{align}
\frac{GH^4\beta}{\pi\gamma}&\,\left[ \frac{1}{\left(2+\beta\right)}\left(\frac{GH^2\beta}{\pi\alpha_-}\right)^{\beta}~
2F_{1}\left(1+\beta, 2+\beta, 3+\beta, -\frac{GH^2\beta}{\pi\alpha_-}\right) \right. \nonumber\\ 
&\, \left. + \frac{1}{\left(2-\beta\right)}\left(\frac{GH^2\beta}{\pi\alpha_+}
\right)^{-\beta}~2F_{1}\left(1-\beta, 2-\beta, 3-\beta, -\frac{GH^2\beta}{\pi\alpha_+}\right) \right] = \frac{8\pi G\rho}{3} + \frac{\Lambda}{3} \,,
\label{FRW-2}
\end{align}
where $\Lambda$ is the integration constant (known as the cosmological constant) and $2F_1(\mathrm{arguments})$ denotes the 
Hypergeometric function. Eq.~(\ref{FRW-1}) and Eq.~(\ref{FRW-2}) represent the modified Friedmann 
equations corresponding to the generalized entropy function $S_\mathrm{g}$. At this stage, we may define the following energy density and pressure as,
\begin{align}
\rho_\mathrm{g} = \frac{3}{8\pi G}\left\{ H^2 - \frac{GH^4\beta}{\pi\gamma}\right. &\,\left[ \frac{1}{\left(2+\beta\right)}
\left(\frac{GH^2\beta}{\pi\alpha_-}\right)^{\beta}~2F_{1}\left(1+\beta, 2+\beta, 3+\beta, -\frac{GH^2\beta}{\pi\alpha_-}\right) \right. \nonumber\\ 
& \left. \left. + \frac{1}{\left(2-\beta\right)}\left(\frac{GH^2\beta}{\pi\alpha_+}\right)^{-\beta}~2F_{1}\left(1-\beta, 2-\beta, 3-\beta, -\frac{GH^2\beta}{\pi\alpha_+}\right) 
\right] \right\} \,,
\label{efective energy density}
\end{align}
and
\begin{align}
p_\mathrm{g} = \frac{\dot{H}}{4\pi G}\left\{\frac{1}{\gamma}\left[\alpha_{+}\left(1 + \frac{\pi \alpha_+}{GH^2\beta}\right)^{\beta - 1} 
+ \alpha_-\left(1 + \frac{\pi \alpha_-}{GH^2\beta}\right)^{-\beta-1}\right] - 1\right\} - \rho_\mathrm{g} \,,
\label{effecyive pressure}
\end{align}
respectively. As a consequence, Eq.~(\ref{FRW-1}) and Eq.~(\ref{FRW-2}) can be equivalently expressed as, 
\begin{align}
\dot{H}=&\,-4\pi G\left[\left(\rho + \rho_\mathrm{g}\right) + \left(p + p_\mathrm{g}\right)\right] \,,\nonumber\\
H^2=&\, \frac{8\pi G}{3}\left(\rho + \rho_\mathrm{g}\right) + \frac{\Lambda}{3} \,.
\label{final FRW}
\end{align}
Therefore due to the above expressions of $\rho_\mathrm{g}$ and $p_\mathrm{g}$, Eq.~(\ref{FRW-1}) and Eq.~(\ref{FRW-2}) 
become similar to the usual Friedmann equations 
where the effective energy density and effective pressure are given by $\rho_\mathrm{eff} = \rho + \rho_\mathrm{g}$ and 
$p_\mathrm{eff} = p + p_\mathrm{g}$. 
Hence $\rho_\mathrm{g}$ and $p_\mathrm{g}$ represent the energy density and pressure corresponding to the entropy function $S_\mathrm{g}$. 
In the next section, we aim to study 
the cosmological implications of the modified Friedmann Eq.~(\ref{final FRW}). In particular, we will show that the Eq.~(\ref{final FRW}) eventually 
leads to a unified scenario of the early inflation to the late dark energy era of the universe.

However, before moving to such a unified description, we will show that the entropic cosmology of the entropy function $S_\mathrm{g}$ can be equivalent 
to the generalized holographic cosmology with a specific holographic cut-off.

\section{Equivalence between the entropic cosmology and the generalized holographic cosmology}\label{sec-equivalence}

In the holographic principle, the holographic energy
density is defined by
\begin{align}
\label{basic}
\rho_\mathrm{hol}=\frac{3c^2}{\kappa^2 L^2_\mathrm{IR}}\, ,
\end{align}
where $L_\mathrm{IR}$ is the infrared cut-off. Here $\kappa^2=8\pi G$ is the gravitational constant and $c$ is a free parameter. 
In this regard, the particle horizon $L_\mathrm{p}$ or the future event horizon
$L_\mathrm{f}$ are given by the following expressions,
\begin{align}
\label{H3}
L_\mathrm{p}\equiv a \int_0^t\frac{dt}{a} \,,\quad
L_\mathrm{f}\equiv a \int_t^\infty \frac{dt}{a}\, .
\end{align}
Taking differentiation of both sides of the above expressions with respect to $t$ 
result to the Hubble parameter in terms of $L_\mathrm{p}$, $\dot{L}_\mathrm{p}$ or in terms of 
$L_\mathrm{f}$, $\dot{L}_\mathrm{f}$ as,
\begin{align}
\label{HLL}
H \left( L_\mathrm{p} , \dot{L}_\mathrm{p} \right) = \frac{\dot{L}_\mathrm{p}}{L_\mathrm{p}} - \frac{1}{L_\mathrm{p}}\, , 
\quad H(L_\mathrm{f} , \dot{L}_\mathrm{f}) = \frac{\dot{L}_\mathrm{f}}{L_\mathrm{f}} + \frac{1}{L_\mathrm{f}} \, .
\end{align}
A general form of the cutoff was proposed in \cite{Nojiri:2005pu} as follows,
\begin{align}
\label{GeneralLIR}
L_\mathrm{IR} = L_\mathrm{IR} \left( L_\mathrm{p}, \dot L_\mathrm{p}, 
\ddot L_\mathrm{p}, \cdots, L_\mathrm{f}, \dot L_\mathrm{f}, \cdots, a\right)\, .
\end{align}
Here it may be mentioned that the other dependency of $L_\mathrm{IR}$, particularly on the Ricci scalar and their derivatives, are captured 
by either $L_\mathrm{p}$ and their derivatives or $L_\mathrm{f}$ and their derivatives via Eq.~(\ref{HLL}).
The above cutoff could be chosen to be equivalent to a general covariant gravity model,
\begin{align}
\label{GeneralAc}
S = \int d^4 \sqrt{-g} F \left( R,R_{\mu\nu} R^{\mu\nu},
R_{\mu\nu\rho\sigma}R^{\mu\nu\rho\sigma}, \Box R, \Box^{-1} R,
\nabla_\mu R \nabla^\mu R, \cdots \right) \, .
\end{align}
Note that generalized holographic DE of \cite{Nojiri:2005pu} gives all known HDEs as specific particular examples of it 
\cite{Nojiri:2021iko,Nojiri:2021jxf}. 
With the help of the generalized cut-off, we will show that the entropic cosmology corresponding to the entropy function of 
Eq.~(\ref{gen-entropy}) can be equivalently mapped to the generalized holographic cosmology 
where the holographic cut-offs are expressed in terms of the $L_\mathrm{p}$ and $\dot{L}_\mathrm{p}$ or 
in terms of the $L_\mathrm{f}$ and $\dot{L}_\mathrm{f}$.

The comparison of Eq.~(\ref{efective energy density}) and Eq.~(\ref{basic}) reveals that the entropic energy density belongs 
from the generalized holographic family, where the corresponding infrared cutoff $L_\mathrm{g}$ is given by, 
\begin{align}
\label{e-1}
\frac{3c^2}{\kappa^2 L^2_\mathrm{g}} 
= \frac{3}{8\pi G} 
\left\{\left( \frac{\dot{L}_\mathrm{p}}{L_\mathrm{p}} - \frac{1}{L_\mathrm{p}} \right)^2 \right. 
&\, -\frac{G\beta}{\pi\gamma}\left( \frac{\dot{L}_\mathrm{p}}{L_\mathrm{p}} - \frac{1}{L_\mathrm{p}} \right)^4
\left[\frac{1}{\left(2+\beta\right)}\left(f_1\left(L_\mathrm{p},\dot{L}_\mathrm{p} \right)
\right)^{\beta}~2F_{1}\left(1+\beta, 2+\beta, 3+\beta, -f_1\left(L_\mathrm{p},\dot{L}_\mathrm{p} \right)\right)
\right. \nonumber\\ 
&\, \left. \left. +\frac{1}{\left(2-\beta\right)}\left(f_2\left(L_\mathrm{p},\dot{L}_\mathrm{p} \right)
\right)^{-\beta}~2F_{1}\left(1-\beta, 2-\beta, 3-\beta, -f_2\left(L_\mathrm{p},\dot{L}_\mathrm{p} \right)\right) \right] \right\} \,,
\end{align}
in terms of $L_\mathrm{p}$ and its derivatives, where $f_1\left(L_\mathrm{p},\dot{L}_\mathrm{p} \right)$ and 
$f_2\left(L_\mathrm{p},\dot{L}_\mathrm{p} \right)$ have the following 
forms:
\begin{align}
f_1\left(L_\mathrm{p},\dot{L}_\mathrm{p} \right)=&\,
\frac{G\beta}{\pi\alpha_-}\left( \frac{\dot{L}_\mathrm{p}}{L_\mathrm{p}} - \frac{1}{L_\mathrm{p}} \right)^2{\color{red} \,,}\nonumber\\
f_2\left(L_\mathrm{p},\dot{L}_\mathrm{p} \right)=&\, \frac{G\beta}{\pi\alpha_+}\left( \frac{\dot{L}_\mathrm{p}}{L_\mathrm{p}} - \frac{1}{L_\mathrm{p}} \right)^2{\color{red} \,,}
\end{align}
respectively. To arrive the above expressions, we use Eq.~(\ref{HLL}). Similarly, $L_\mathrm{g}$ in terms of the future horizon and its 
derivatives turns out to be,
\begin{align}
\label{e-2}
\frac{3c^2}{\kappa^2 L^2_\mathrm{g}} 
= \frac{3}{8\pi G} \left\{\left(\frac{\dot{L}_\mathrm{f}}{L_\mathrm{f}} +\frac{1}{L_\mathrm{f}} \right)^2 \right.
-&\, \frac{G\beta}{\pi\gamma}\left( \frac{\dot{L}_\mathrm{f}}{L_\mathrm{f}} + \frac{1}{L_\mathrm{f}} \right)^4
\left[\frac{1}{\left(2+\beta\right)}\left(h_1\left(L_\mathrm{f},\dot{L}_\mathrm{f} \right)
\right)^{\beta}~2F_{1}\left(1+\beta, 2+\beta, 3+\beta, -h_1\left(L_\mathrm{f},\dot{L}_\mathrm{f} \right)\right) \right. \nonumber\\ 
+&\, \frac{1}{\left(2-\beta\right)}\left(h_2\left(L_\mathrm{f},\dot{L}_\mathrm{f} \right)\right)^{-\beta}~ 2F_{1}
\left(1-\beta, 2-\beta, 3-\beta, -h_2\left(L_\mathrm{f},\dot{L}_\mathrm{f} \right)\right)\bigg]\bigg\}{\color{red} \,,}
\end{align}
with $h_1\left(L_\mathrm{f},\dot{L}_\mathrm{f} \right)$ and $h_2\left(L_\mathrm{f},\dot{L}_\mathrm{f} \right)$ are given by, 
\begin{align}
h_1\left(L_\mathrm{f},\dot{L}_\mathrm{f} \right)=&\,\frac{G\beta}{\pi\alpha_-}\left( \frac{\dot{L}_\mathrm{f}}
{L_\mathrm{f}} + \frac{1}{L_\mathrm{f}} \right)^2 \,,\nonumber\\
 h_2\left(L_\mathrm{f},\dot{L}_\mathrm{f} \right)=&\,\frac{G\beta}{\pi\alpha_+}\left( \frac{\dot{L}_\mathrm{f}}
 {L_\mathrm{f}} + \frac{1}{L_\mathrm{f}} \right)^2 \,.
\end{align}

Here we would like to determine the EoS parameter of the holographic energy density corresponds to the cut-off $L_\mathrm{g}$, 
in particular of $\rho_\mathrm{hol} = 3c^2/\left(\kappa^2L_\mathrm{g}^2\right)$. 
In this regard, the conservation equation of $\rho_\mathrm{hol}$ immediately yields the respective EoS parameter (symbolized by $\Omega_\mathrm{hol}^{(g)}$) as, 
\begin{align}
\Omega_\mathrm{hol}^{(g)} = -1 - \left(\frac{2}{3HL_\mathrm{g}}\right)\frac{dL_\mathrm{g}}{dt} \,,
\label{eos-holg}
\end{align}
where $L_\mathrm{g}$ is obtained in Eq.~(\ref{e-1}) (or in Eq.~(\ref{e-2})) and the superscript `$\mathrm{g}$' 
in the above expression denotes the EoS parameter corresponds to the holographic cut-off $L_\mathrm{g}$. 
Due to Eq.~(\ref{HLL}), the above form of $\Omega_\mathrm{hol}^{(g)}$ seems to be 
equivalent to the EoS of the entropic energy density given by $\omega_\mathrm{g} = p_\mathrm{g}/\rho_\mathrm{g}$, i.e.,  
\begin{align}
\Omega_\mathrm{hol}^{(g)} \equiv \omega_{g} \,.
\label{e-3}
\end{align}
Therefore one may argue that the cosmology corresponding to the entropy function $S_\mathrm{g}$ of Eq.~(\ref{gen-entropy}) 
is equivalent to the generalized 
holographic cosmology with the holographic cut-off being represented in terms of $L_\mathrm{p}$ and $\dot{L}_\mathrm{p}$ (see 
Eq.~(\ref{e-1})) or in  terms of $L_\mathrm{f}$ and $\dot{L}_\mathrm{f}$ (see Eq.~(\ref{e-2})). 

\section{Early universe cosmology from the generalized entropy function}\label{sec-inf}

During the early stage of the universe we consider the matter field and the cosmological constant ($\Lambda$) to 
be absent, i.e., $\rho = p = \Lambda = 0$. During the early universe, the cosmological constant is highly suppressed with respect to the 
entropic energy density and thus we can safely neglect the $\Lambda$ in studying the early inflationary scenario of the universe. Actually 
we will show that the entropic energy density will be the main agent to trigger the inflation. 
Here we would like to mention that although the cosmological constant does not play any role in early-time universe, it may contribute 
a significant role in the late dark energy epoch. We will demonstrate this in Sec.[\ref{sec-DE}].

In the early-time universe when $\rho = p = \Lambda = 0$, Eq.~(\ref{FRW-1}) and 
Eq.~(\ref{FRW-2}) become,
\begin{align}
\left[\alpha_{+}\left(1 + \frac{\pi \alpha_+}{\beta GH^2}\right)^{\beta - 1} 
+ \alpha_-\left(1 + \frac{\pi \alpha_-}{\beta GH^2}\right)^{-\beta-1}\right]\dot{H} = 0 \,,
\label{FRW-1-inf}
\end{align}
and
\begin{align}
& \left[\frac{1}{\left(2+\beta\right)}\left(\frac{GH^2\beta}{\pi\alpha_-}\right)^{\beta}~2F_{1}
\left(1+\beta, 2+\beta, 3+\beta, -\frac{GH^2\beta}{\pi\alpha_-}\right) \right. \nonumber\\ 
&\left. \quad +\frac{1}{\left(2-\beta\right)}\left(\frac{GH^2\beta}{\pi\alpha_+}\right)^{-\beta}~2F_{1}
\left(1-\beta, 2-\beta, 3-\beta, -\frac{GH^2\beta}{\pi\alpha_+}\right) \right] = 0 \, ,
 \label{FRW-2-inf}
\end{align}
respectively. Since $\alpha_+$ and $\alpha_-$ are positive, the solution of Eq.~(\ref{FRW-1-inf}) is given by $\dot{H} = 0$ or $H = H_0$ with 
$H_0$ being a constant which can be determined by solving the other Friedmann Eq.~(\ref{FRW-2-inf}). 
The solution of a constant Hubble parameter (in absence of matter field) is a generic feature of entropic cosmology 
(see \cite{Nojiri:2021iko} 
for the Tsallis entropy or the R\'{e}nyi entropy). Here it may be mentioned that the typical 
energy scale during early universe is of the order $\sim 10^{16}\mathrm{GeV}$ ($= 10^{-3}M_\mathrm{Pl}$ where recall that $M_\mathrm{Pl}$ is the Planck mass 
and $M_\mathrm{Pl} = 1/\sqrt{16\pi G}$). This indicates that the condition $GH^2 \ll 1$ holds during the early phase of the universe. 
Owing to such condition, we can safely expand the Hypergeometric function of Eq.~(\ref{FRW-2-inf}) 
as the Taylor series with respect to the argument containing $GH^2$. 
As a result, Eq.~(\ref{FRW-2-inf}) can be expressed as,
\begin{align}
\frac{1}{2-\beta} - \left(\frac{1-\beta}{3-\beta}\right)\frac{GH^2\beta}{\pi \alpha_+} = 0 \,,
\label{dS Hubble parameter-eq}
\end{align}
where we keep the terms up-to-the leading order in $GH^2$. 
The above equation immediately provides the de-Sitter Hubble parameter as,
\begin{align}
H = 4\pi M_\mathrm{Pl}\sqrt{\frac{\alpha_+}{\beta}}\left[\frac{(3-\beta)}{(2-\beta)(1-\beta)}\right] \,.
\label{dS Hubble parameter-sol}
\end{align}
For $\frac{\alpha_+}{\beta} \sim 10^{-6}$ and $\beta \lesssim \mathcal{O}(1)$, the constant Hubble parameter can be fixed at 
$H \sim 10^{-3}M_\mathrm{Pl}$ which can be identified with typical inflationary energy scale. 
Therefore the entropic cosmology corresponding to the 
generalized entropy function $S_\mathrm{g}$ leads to a de-Sitter inflationary scenario during the early universe. 
However, a de-Sitter inflation has no exit mechanism, i.e.,  the inflation becomes eternal; and moreover 
the scalar spectral index of primordial curvature perturbation gets exactly 
scale invariant in the context of a de-Sitter inflation, which is not consistent with the recent Planck data \cite{Planck:2018jri} at all. 
This indicates that the constant 
Hubble parameter obtained in Eq.~(\ref{dS Hubble parameter-sol}) does not lead to a good inflationary scenario of the universe.  

\subsection{Viable inflation}

It is well known that a quasi de-Sitter evolution of the Hubble parameter may result in viable inflation. 
Therefore in the present context, we consider the parameters of $S_\mathrm{g}$ to be slowly varying functions 
with respect to the cosmic time (see \cite{Nojiri:2019skr}, for a case where the Tsallis entropy with $varying$ exponent was studied). 
In particular, we consider the parameter 
$\gamma$ to vary and the other parameters (i.e., $\alpha_+$, $\alpha_-$ and $\beta$) remain constant with $t$. 
The running behavior of $\gamma$ 
may be motivated by quantum gravity, particularly in the case of gravity, if the space-time fluctuates at high energy scales, 
the degrees of freedom may increase. On the other hand, if gravity becomes a topological theory, the degrees of freedom may decrease. 
We consider, 
\begin{align}
\gamma(N)=\left\{ 
\begin{array}{ll}
\gamma_0~\exp{\left[-\int_{N}^{N_f}\sigma(N)~dN\right]}\quad &;\ N \leq N_f \\
\gamma_0 &;\ N \geq N_f \,,
\end{array}
\right. 
\label{gamma function}
\end{align}
where $\gamma_0$ is a constant and $N$ denotes the inflationary e-folding number with $N_f$ being the total e-folding number of the inflationary era. 
The e-folding number up-to the time $t$ is defined as $N= \int^{t}Hdt$, where $N = 0$ refers the instance of the beginning of inflation which, 
in the present context, we consider to happen when the CMB scale mode ($\sim 0.05\mathrm{Mpc}^{-1}$) crosses the Hubble horizon. 
The function $\sigma(N)$ has the following form,
\begin{align}
\sigma(N) = \sigma_0 + \mathrm{e}^{-\left(N_f - N\right)} \,,
\label{sigma function}
\end{align}
where $\sigma_0$ is a constant. 
The second term in the expression of $\sigma(N)$ becomes effective only when $N\approx N_f$, i.e.,  near the end of inflation. 
The term $\mathrm{e}^{-\left(N_f - N\right)}$ in Eq.~(\ref{sigma function}) is actually considered to ensure an exit 
from inflation era and thus proves 
to be an useful one to make the inflationary scenario viable.\\

\begin{figure}[!h]
\begin{center}
\centering
\includegraphics[width=3.5in,height=2.5in]{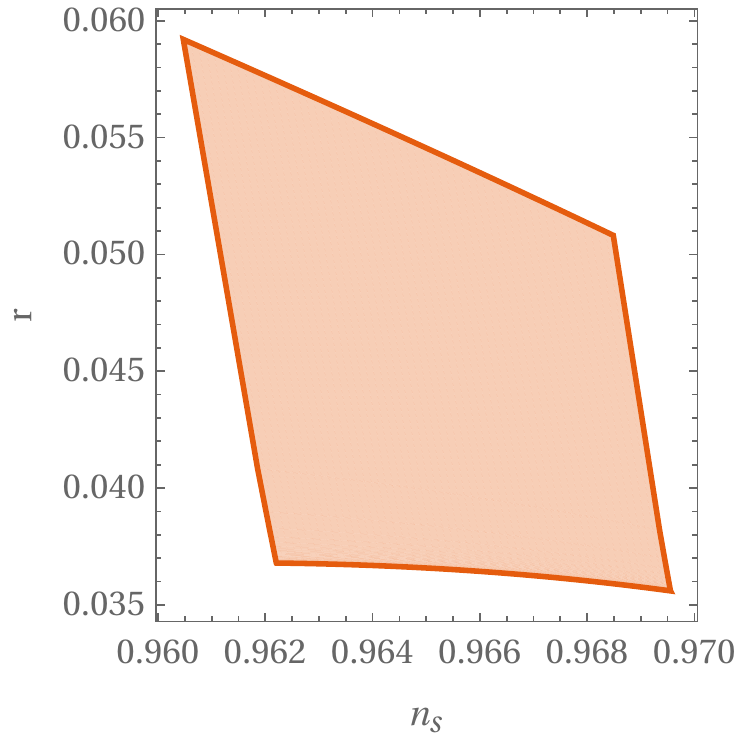}
\caption{Simultaneous compatibility of $n_s$ and $r$ according to Eq.~(\ref{ns final form}) and Eq.~(\ref{r final form}). This corresponds 
 to the generalized entropy function $S_\mathrm{g}$ where $\gamma$ varies according to 
Eq.~(\ref{gamma function}). In the plot, we take $N_f = 58$, $\sigma_0 = [0.013,0.017]$ and $\left(\alpha_+/\alpha_-\right)^{\beta} = [7.5,10^4]$ 
respectively.} 
 \label{plot-observable}
\end{center}
\end{figure}

Thus as a whole, in the case where $\gamma$ varies with $N$, the generalized entropy function looks like the expression of Eq.~(\ref{gen-entropy}) 
with $\gamma$ replaced by $\gamma(N)$ shown in Eq.~(\ref{gamma function}). 
Consequently, the modified Friedmann equation corresponding to the entropy function $S_\mathrm{g}$ comes as,
\begin{align}
 -\left(\frac{2\pi}{G}\right)\left(\frac{\partial S_\mathrm{g}}{\partial S}\right)\frac{H'(N)}{H^3} + 
\left(\frac{\partial S_\mathrm{g}}{\partial \gamma}\right)\gamma'(N) = 0 \,,
\label{FRW-eq-viable-inf-sub}
\end{align}
where we consider $\rho = p = 0$ to describe the early universe, and an overprime denotes $\frac{d}{dN}$. 
Using the explicit form of $S_\mathrm{g}$, the above equation takes the following form,
\begin{align}
 -\left(\frac{2\pi}{G}\right)
\left[\frac{\alpha_+\left(1 + \frac{\alpha_+}{\beta}~S\right)^{\beta-1} + \alpha_-\left(1 + \frac{\alpha_-}{\beta}~S\right)^{-\beta-1}}
{\left(1 + \frac{\alpha_+}{\beta}~S\right)^{\beta} - \left(1 + \frac{\alpha_-}{\beta}~S\right)^{-\beta}}\right]\frac{H'(N)}{H^3} = 
\frac{\gamma'(N)}{\gamma} \,.
\label{FRW-eq-viable-inf-main}
\end{align}
Eq.~(\ref{FRW-eq-viable-inf-main}) clearly indicates that due to $\gamma'(N)\neq 0$, the Hubble parameter is not a constant in this context, 
and we may get a quasi de-Sitter inflationary scenario when the $\gamma$ varies with the e-folding number. 
From Eq.~(\ref{gamma function}), we obtain,
\begin{align}
\gamma'(N)/\gamma(N) = \sigma(N) \,.
\label{gamma function derivative-1}
\end{align}
during the early stage of the universe. Consequently, Eq.~(\ref{FRW-eq-viable-inf-main}) becomes, 
\begin{align}
 -\left(\frac{2\pi}{G}\right)
\left[\frac{\alpha_+\left(1 + \frac{\alpha_+}{\beta}~S\right)^{\beta-1} + \alpha_-\left(1 + \frac{\alpha_-}{\beta}~S\right)^{-\beta-1}}
{\left(1 + \frac{\alpha_+}{\beta}~S\right)^{\beta} - \left(1 + \frac{\alpha_-}{\beta}~S\right)^{-\beta}}\right]\frac{H'(N)}{H^3} = \sigma(N) \,.
\label{FRW-eq-viable-inf-main1}
\end{align}
Recall that $S = \pi/(GH^2)$, from which, we get $2HdH = -\frac{\pi}{GS^2}dS$. 
Using these relations, Eq.~(\ref{FRW-eq-viable-inf-main1}) can be equivalently written as,
\begin{align}
\left[\frac{\alpha_+\left(1 + \frac{\alpha_+}{\beta}~S\right)^{\beta-1} + \alpha_-\left(1 + \frac{\alpha_-}{\beta}~S\right)^{-\beta-1}}
{\left(1 + \frac{\alpha_+}{\beta}~S\right)^{\beta} - \left(1 + \frac{\alpha_-}{\beta}~S\right)^{-\beta}}\right]dS = \sigma(N)dN \,,
\label{FRW-eq-viable-inf-main2}
\end{align}
on integrating which, we obtain the solution of $H = H(N)$ as follows,
\begin{align}
\left(1 + \frac{\pi \alpha_+}{GH^2\beta}\right)^{\beta} - \left(1 + \frac{\pi\alpha_-}{GH^2\beta}\right)^{-\beta} 
= \exp{\left[\int_0^{N}\sigma(N)dN\right]}
\label{solution-viable-inf-1}
\end{align}
with $\int_0^{N}\sigma(N)dN = N\sigma_0 + \mathrm{e}^{-(N_f - N)} - \mathrm{e}^{-N_f}$ from Eq.~(\ref{sigma function}). 
Due to the condition $GH^2 \ll 1$ during the early universe (see the discussion after Eq.~(\ref{FRW-2-inf})), 
the Hubble parameter from Eq.~(\ref{solution-viable-inf-1}) can be solved as,
\begin{align}
H(N) = 4\pi M_\mathrm{Pl}\sqrt{\frac{\alpha_+}{\beta}}
\left[\frac{2^{1/(2\beta)}\exp{\left[-\frac{1}{2\beta}\int_0^{N}\sigma(N)dN\right]}}
{\left\{1 + \sqrt{1 + 4\left(\alpha_+/\alpha_-\right)^{\beta}\exp{\left[-2\int_0^{N}\sigma(N)dN\right]}}\right\}^{1/(2\beta)}}\right] \,,
\label{solution-viable-inf-2}
\end{align}
where $M_\mathrm{Pl} = 1/\sqrt{16\pi G}$. Eq.~(\ref{solution-viable-inf-2}) represents the evolution of the Hubble parameter where $\gamma(N)$ varies 
according to Eq.~(\ref{gamma function}) and the other parameters of $S_\mathrm{g}$ are considered to be constant. 
In order to examine whether the above solution of $H = H(N)$ leads to an accelerating stage of the universe, we calculate the slow roll parameter 
(defined by $\epsilon = -H'(N)/H(N)$) as,
\begin{align}
\epsilon(N) = \frac{\sigma(N)}
{2\beta\sqrt{1 + 4\left(\alpha_+/\alpha_-\right)^{\beta}\exp{\left[-2\int_0^{N}\sigma(N)dN\right]}}} \,.
\label{slow roll parameter}
\end{align}
Furthermore we calculate $\epsilon'(N)$ (which is important to determine various observable indices, as we will show later) as,
\begin{align}
\epsilon'(N)/\epsilon(N) = \frac{\sigma(N)}
{1 + \frac{1}{4}\left(\alpha_+/\alpha_-\right)^{-\beta}\exp{\left[2\int_0^{N}\sigma(N)dN\right]}} 
+ \frac{\mathrm{e}^{-\left(N_f - N\right)}}{\sigma(N)} \,.
\label{derivative of slow roll parameter}
\end{align}
Since $\sigma(N) > 0$, it is clear that $\epsilon(N)$ is an increasing function of $N$. 
Therefore we can consider the parameter values in such a way that $\epsilon(N)$ becomes less than unity for $N < N_f$ and $\epsilon(N_f) = 1$. 
The condition $\epsilon(N) < 1$ ensures the accelerating stage of the universe and $\epsilon(N_f) = 1$ indicates the end of inflation at $N = N_f$. 
However before scanning the parameter values, we first calculate various observable 
indices like the spectral index for primordial curvature perturbation ($n_s$) and the tensor-to-scalar ratio ($r$). 
They are defined by, 
\begin{align}
n_s = 1 - 2\epsilon - 2\epsilon'/\epsilon\bigg|_ \mathrm{h.c} \,, \quad 
r = 16\epsilon\bigg|_\mathrm{h.c} \,,
\nonumber
\end{align}
where the suffix `h.c' denotes the horizon crossing instant of the large scale CMB mode, which actually refers to the beginning of inflation, i.e.,  
$N = 0$ (see the discussion after Eq.~(\ref{gamma function})). 
Plugging back the expressions of $\epsilon(N)$ and $\epsilon'(N)$ into the above equation, 
we obtain,
\begin{align}
n_s=&\,1 - \frac{\sigma_0}{\beta\sqrt{1 + 4\left(\alpha_+/\alpha_-\right)^{\beta}}} - \frac{8\sigma_0\left(\alpha_+/\alpha_-\right)^{\beta}}
{1 + 4\left(\alpha_+/\alpha_-\right)^{\beta}} \,,\nonumber\\
r=\,&\frac{8\sigma_0}{\beta\sqrt{1 + 4\left(\alpha_+/\alpha_-\right)^{\beta}}}
\label{ns and r}
\end{align}
where we neglect the term $\mathrm{e}^{-N_f} \ll 1$ (as $N_f \approx 55$ or $58$). 
Moreover in order to ensure the end of inflation at $N= N_f$, we put 
$\epsilon(N_f) = 1$ which, due to Eq.~(\ref{slow roll parameter}), leads to the following relation between the parameters,
\begin{align}
\beta = \frac{(1 + \sigma_0)}
{2\sqrt{1 + 4\left(\alpha_+/\alpha_-\right)^{\beta}\exp{\left[-2\left(1 + \sigma_0N_f\right)\right]}}} \,,
\label{end of inflation}
\end{align}
where we use $\int_0^{N_f}\sigma(N)dN = 1+\sigma_0N_f$. 
Using the above expression of $\beta$ into Eq.~(\ref{ns and r}), we get the final forms of the scalar spectral index and the tensor-to-scalar ratio as, 
\begin{align}
n_s = 1 - \frac{2\sigma_0\sqrt{1 + 4\left(\alpha_+/\alpha_-\right)^{\beta}\exp{\left[-2\left(1 + \sigma_0N_f\right)\right]}}}
{(1+\sigma_0)\sqrt{1 + 4\left(\alpha_+/\alpha_-\right)^{\beta}}} - \frac{8\sigma_0\left(\alpha_+/\alpha_-\right)^{\beta}}
{1 + 4\left(\alpha_+/\alpha_-\right)^{\beta}} \,,
\label{ns final form}
\end{align}
and 
\begin{align}
r = \frac{16\sigma_0\sqrt{1 + 4\left(\alpha_+/\alpha_-\right)^{\beta}\exp{\left[-2\left(1 + \sigma_0N_f\right)\right]}}}
{(1+\sigma_0)\sqrt{1 + 4\left(\alpha_+/\alpha_-\right)^{\beta}}}
\label{r final form}
\end{align}
respectively, which indeed contain the information of $\epsilon(N_f) = 1$. 
It may be noticed that $n_s$ and $r$ depend on the dimensionless parameters $\sigma_0$ and $\left(\alpha_+/\alpha_-\right)^{\beta}$. 
We can now directly confront the spectral index and the tensor-to-scalar ratio with the Planck 2018 constraints \cite{Planck:2018jri}, 
which constrain the observational indices as follows, 
\begin{align}
n_s = 0.9649 \pm 0.0042 \,, \quad r < 0.064 \,.\nonumber
\end{align}
For the model at hand, the theoretical expectations of 
$n_s$ and $r$ get simultaneously compatible with the Planck constraints for the following ranges of parameter values: 
$\sigma_0 = [0.013,0.017]$ and $\left(\alpha_+/\alpha_-\right)^{\beta} \geq 7.5$ respectively, where we consider $N_f = 58$. 
Here we would like to mention that at large $\left(\alpha_+/\alpha_-\right)^{\beta}$, both the spectral index and the tensor-to-scalar ratio become 
independent of $\left(\alpha_+/\alpha_-\right)^{\beta}$ and thus the viable range of $\left(\alpha_+/\alpha_-\right)^{\beta}$ seems to be 
unbounded from above. The simultaneous compatibility of $n_s$ and $r$ is depicted in Fig.~\ref{plot-observable}. 
Using Eq.~(\ref{end of inflation}) along with the aforementioned ranges of $\sigma_0$ and 
$\left(\alpha_+/\alpha_-\right)^{\beta}$, we estimate the parameter $\beta$ which turns out to lie within a small range, in particular, 
we get: $0 < \beta \leq 0.4$. 
Therefore, in order to have a viable inflationary scenario during the early universe, the parameters of $S_\mathrm{g}$ should lie 
within the following ranges:
\begin{align}
\sigma_0=&\,[0.013,0.017] \,, \quad \left(\alpha_+/\alpha_-\right)^{\beta} \geq 7.5 \,,\nonumber\\
\beta=&\, (0,0.4]\ \mbox{and}\ \left(\alpha_+/\beta\right) \approx 10^{-6} \,,
\label{inf-constraints}
\end{align}
for $N_f = 58$. The consideration of $\frac{\alpha_+}{\beta} \sim 10^{-6}$ leads to the energy scale at the onset 
of inflation as $H \sim 10^{-3}M_\mathrm{Pl}$.

From the Fig.~\ref{plot-observable}, we notice that the smallest value of the tensor-to-scalar ratio is $r \approx 0.035$. This can be 
demonstrated from Eq.(\ref{r final form}) which indicates that $r$ depends on $\sigma_0$ and $\left(\alpha_+/\alpha_-\right)^{\beta}$, and moreover, 
the value of $r$ decreases with the increasing value of $\left(\alpha_+/\alpha_-\right)^{\beta}$. Therefore the smallest value of $r$ occurs at the limit of 
$\left(\frac{\alpha_+}{\alpha_-}\right)^{\beta}\rightarrow \infty$, and at this limit, 
Eq.(\ref{r final form}) can be written as,
\begin{align}
r = \frac{16\sigma_0~\exp{\left[-\left(1 + \sigma_0N_f\right)\right]}}
{(1+\sigma_0)}~~,
\label{r approximated form}
\end{align}
which turns out to be independent of $\left(\alpha_+/\alpha_-\right)^{\beta}$ and depends only on $\sigma_0$. As a result, 
due to the viable range of $\sigma_0 = [0.013,0.017]$ (see Eq.(\ref{inf-constraints})) along with $N_f = 58$, 
the smallest value of the tensor-to-scalar ratio from Eq.(\ref{r approximated form}) becomes $r \approx 0.035$, 
as we show in the Fig.~\ref{plot-observable}. Here it may be mentioned that the tensor-to-scalar ratio may get smaller value than 
$r = 0.035$ in some holographic inflationary model(s) (see \cite{Nojiri:2019kkp} for instance).\\

\begin{figure}[!h]
\begin{center}
\centering
\includegraphics[width=3.5in,height=2.5in]{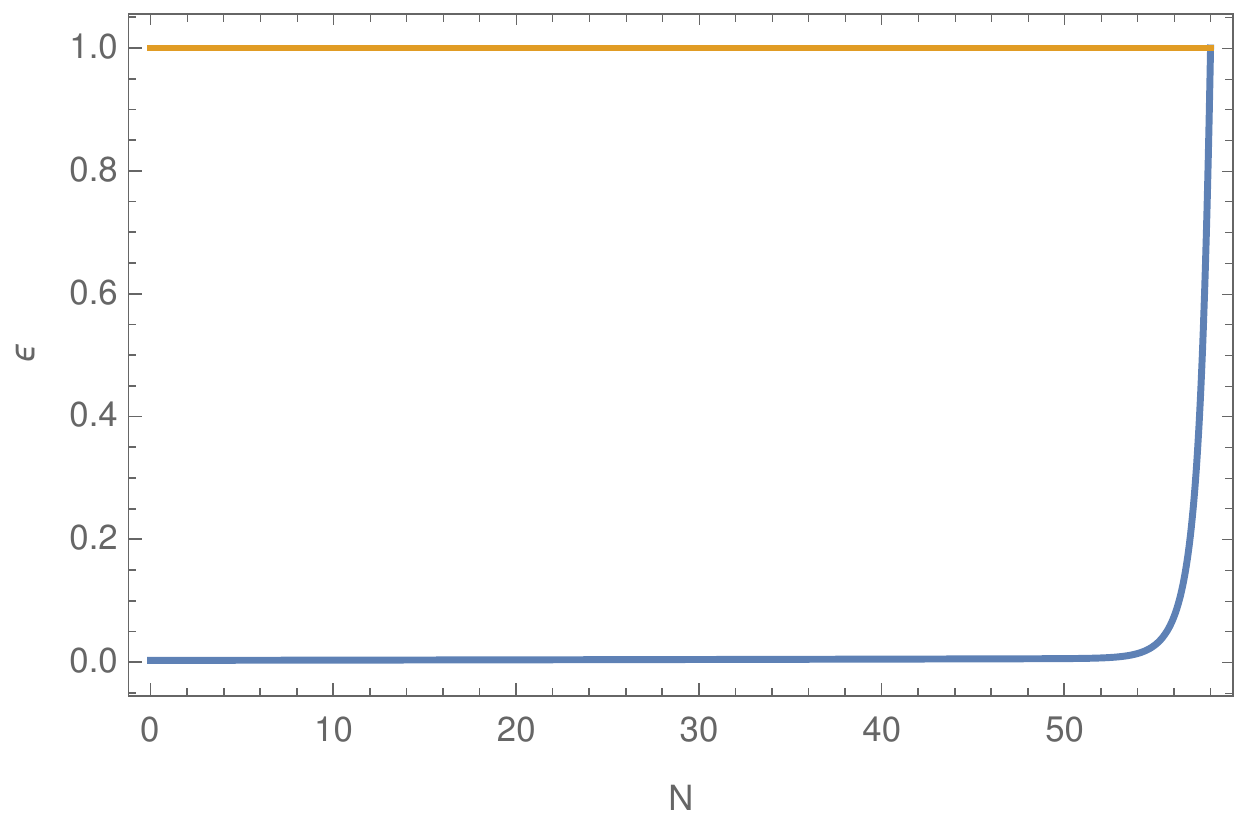}
\caption{$\epsilon(N)$ vs. $N$ from Eq.~(\ref{slow roll parameter}). Here we take 
$\sigma_0 = 0.014$, $\left(\alpha_+/\alpha_-\right)^{\beta} = 10$ and $N_f = 58$.} 
\label{plot-sr}
\end{center}
\end{figure}

Taking a particular set of values of the parameters from their viable ranges mentioned in Eq.(\ref{inf-constraints}), say 
$N_f = 58$, $\sigma_0 = 0.014$ and $\left(\alpha_+/\alpha_-\right)^{\beta} = 10$, 
we give the plot 
of $\epsilon(N)$ vs. $N$ from Eq.~(\ref{slow roll parameter}), see the Fig.~\ref{plot-sr}. 
The figure clearly suggests that -- (1) $\epsilon(N)$ is an increasing function 
with respect to the e-folding number during $N \leq N_f = 58$, (2) $\epsilon(N)$ remains less than unity (and almost constant) 
for $N < N_f = 58$ and rises to unity at $N = N_f$. This ensures that the universe experiences an inflationary era having an exit point, and 
the total inflationary e-folding number is given by $N_f = 58$. 

Here we would like to mention that if one considers a larger inflationary e-folding number, say $N_f = 62$, then the viable ranges of the parameters 
are given by : $\sigma_0 = [0.0133,0.0175]$ and $\left(\frac{\alpha_+}{\alpha_-}\right)^{\beta} \geq 7.5$ in order to get a simultaneous compatibility 
of the scalar spectral index and the tensor-to-scalar ratio. Moreover, the smallest value of tensor-to-scalar ratio for $N_f = 62$ comes as 
$r \approx 0.034$. Therefore the viable ranges of the parameters do not change drastically for two different values of $N_f$, however the smallest value 
of $r$ gets reduced for larger inflationary e-fold number.\\

Thus as a whole, the entropic cosmology corresponding to the generalized entropy function $S_\mathrm{g}$ where the parameter $\gamma$ is a slowly varying function 
of $N$ according to Eq.~(\ref{gamma function}), triggers a viable inflationary scenario during the early universe. 
In particular -- (1) the inflation has an exit mechanism at around 58 e-folding number, which is consistent with the resolution of 
horizon and flatness problems, and (2) the scalar spectral index for curvature perturbation and the tensor-to-scalar 
ratio turn out to be simultaneously compatible with the recent Planck data for suitable values of the parameters obtained in Eq.~(\ref{inf-constraints}).

\section{Dark energy era from the generalized entropy function}\label{sec-DE}

In this section we will concentrate on late time cosmological implications of the generalized entropy function ($S_\mathrm{g}$). 
During the late time, the parameter $\gamma$ becomes constant, in particular $\gamma = \gamma_0$, as we demonstrated in Eq.~(\ref{gamma function}). 
As a result, the entropy function at the late time takes the following form,
\begin{align}
S_\mathrm{g} = \frac{1}{\gamma_0}\left[\left(1 + \frac{\alpha_+}{\beta}~S\right)^{\beta} - \left(1 + \frac{\alpha_-}{\beta}~
S\right)^{-\beta}\right] \,,
\label{entropy-late time}
\end{align}
with $S = \pi/(GH^2)$. 
Consequently, the energy density and pressure corresponding to the $S_\mathrm{g}$ are given by Eq.~(\ref{efective energy density}) 
and Eq.~(\ref{effecyive pressure}) respectively, with $\gamma = \gamma_0$, i.e.,
\begin{align}
 \rho_\mathrm{g} = \frac{3}{8\pi G}\left\{H^2 - 
 \frac{GH^4\beta}{\pi\gamma_0} \right. & \left[ \frac{1}{\left(2+\beta\right)}\left(\frac{GH^2\beta}{\pi\alpha_-}
 \right)^{\beta}~ 2F_{1}\left(1+\beta, 2+\beta, 3+\beta, -\frac{GH^2\beta}{\pi\alpha_-}\right) \right. \nonumber\\ 
&\left. \left. +\frac{1}{\left(2-\beta\right)}\left(\frac{GH^2\beta}{\pi\alpha_+}
\right)^{-\beta}~ 2F_{1}\left(1-\beta, 2-\beta, 3-\beta, -\frac{GH^2\beta}{\pi\alpha_+}\right) \right] \right\} \,,
 \label{efective energy density-late time}
\end{align}
and
\begin{align}
p_\mathrm{g} = \frac{\dot{H}}{4\pi G}\left\{\frac{1}{\gamma_0}\left[\alpha_{+}\left(1 + \frac{\pi \alpha_+}{GH^2\beta}\right)^{\beta - 1} 
+ \alpha_-\left(1 + \frac{\pi \alpha_-}{GH^2\beta}\right)^{-\beta-1}\right] - 1\right\} - \rho_\mathrm{g} \,.
\label{effecyive pressure-late time}
\end{align}
At the present epoch, the typical estimation of the Hubble parameter is of the order $\sim 10^{-40}\mathrm{GeV}$, and thus the condition 
$GH^2 \ll 1$ is safely satisfied during the late stage of the universe. 
Owing to such condition, the above expressions of $\rho_\mathrm{g}$ and $p_\mathrm{g}$ can be approximated by,
\begin{align}
\rho_\mathrm{g}=&\, \frac{3H^2}{8\pi G}\left[1 - \frac{\alpha_+}{\gamma_0(2-\beta)}\left(\frac{GH^2\beta}{\pi\alpha_+}\right)^{1-\beta}\right] \,,
\nonumber\\
p_\mathrm{g}=&\, -\frac{\dot{H}}{4\pi G}\left[1 - \frac{\alpha_+}{\gamma_0}\left(\frac{GH^2\beta}{\pi\alpha_+}\right)^{1-\beta} 
 - \left(\frac{\alpha_+}{\gamma_0}\right)\left(\frac{\alpha_+}{\alpha_-}\right)^{\beta}\left(\frac{GH^2\beta}{\pi\alpha_+}\right)^{1+\beta}\right] 
 - \rho_\mathrm{g} \,.
\label{energy and pressure-late time}
\end{align}
In regard to the dark energy era, we will investigate two distinct cases, 
namely when the explicit cosmological constant $\Lambda$ is absent and when it is present. 

\subsection*{Without the cosmological constant: $\Lambda = 0$}
In this case, the dark energy density ($\rho_\mathrm{D}$) is solely contributed from the entropic energy density, i.e the dark energy density 
and the corresponding pressure are given by,
\begin{eqnarray}
 \rho_\mathrm{D} = \rho_\mathrm{g}~~~~~~~\mathrm{and}~~~~~~~~p_\mathrm{D} = p_\mathrm{g}
\end{eqnarray}
respectively, with $\rho_\mathrm{g}$ and $p_\mathrm{g}$ are shown in Eq.(\ref{energy and pressure-late time}). 
Therefore the dark energy equation of state (EoS) parameter(symbolized by $\omega_\mathrm{g}$) is determined as,
\begin{align}
\omega_\mathrm{g} = p_\mathrm{g}/\rho_\mathrm{g} = -1 - \left(\frac{2\dot{H}}{3H^2}\right)
\left[\frac{1 - \frac{\alpha_+}{\gamma_0}\left(\frac{GH^2\beta}{\pi\alpha_+}\right)^{1-\beta} 
 - \left(\frac{\alpha_+}{\gamma_0}\right)\left(\frac{\alpha_+}{\alpha_-}\right)^{\beta}\left(\frac{GH^2\beta}{\pi\alpha_+}\right)^{1+\beta}}
{1 - \frac{\alpha_+}{\gamma_0(2-\beta)}\left(\frac{GH^2\beta}{\pi\alpha_+}\right)^{1-\beta}}\right] \,.
\label{eos}
\end{align}
Now we aim to examine whether the above form of the EoS parameter, sourced from the generalized entropy function, leads to a viable 
dark energy (DE) era at the current universe. 

We consider that the universe is filled with pressureless dust, as well as with the above entropic energy density. 
Then the two Friedmann equations are written as,
\begin{align}
H^2 = \frac{8\pi G}{3}\left(\rho_m + \rho_\mathrm{g}\right) \,,\nonumber\\
\dot{H} = -4\pi G\left[\rho_m + \left(\rho_\mathrm{g} + p_\mathrm{g}\right)\right] \,,
\label{FRW-late time}
\end{align}
where $\rho_\mathrm{g}$ and $p_\mathrm{g}$ are shown in Eq.~(\ref{energy and pressure-late time}), and $\rho_m$ 
represents the energy density of the dust fluid. 
As usual we consider the two sector to be non-interacting, and thus the usual conservation equations hold in the individual sector, namely
\begin{align}
\dot{\rho}_m + 3H\rho_m = 0 \,,\nonumber\\
\dot{\rho}_\mathrm{g} + 3H\rho_\mathrm{g}\left(1 + \omega_\mathrm{g}\right) = 0 \,.
\label{conservation relation}
\end{align}
Finally we introduce the fractional energy density as, 
\begin{align}
\Omega_m = \left(\frac{8\pi G}{3H^2}\right)\rho_m \,, \quad 
\Omega_\mathrm{g} = \left(\frac{8\pi G}{3H^2}\right)\rho_\mathrm{g} \,,
\label{fractional energy density}
\end{align}
respectively. From Eq.~(\ref{conservation relation}), we obtain $\rho_m = \rho_{m0}\left(\frac{a_0}{a}\right)^3$ with $\rho_{m0}$ being 
the present matter energy density, namely at $a_0$ (in the following, the subscript``0'' with a quantity 
denotes the value of the respective quantity at present). 
Substituting such evolution of $\rho_m$ into Eq.~(\ref{fractional energy density}) leads to 
$\Omega_m = \Omega_{m0}a_0^3H_0^2/(a^3H^2)$ which along with the relation $\Omega_m + \Omega_\mathrm{g} = 1$ results to the Hubble parameter in terms of 
the red shift factor ($z$) as follows,
\begin{align}
H(z) = \frac{H_0\sqrt{\Omega_{m0}(1+z)^3}}{\sqrt{1-\Omega_\mathrm{g}}} \,,
\label{Hubble late time}
\end{align}
where $z = \frac{a_0}{a} - 1$ is the red shift factor (the present time is designated by $z = 0$), and $\Omega_{m0} \approx 0.3153$. 
Plugging the expression of $\rho_\mathrm{g}$ from Eq.~(\ref{energy and pressure-late time}) into 
Eq.~(\ref{fractional energy density}) and using the above form of $H(z)$, we obtain,
\begin{align}
\Omega_\mathrm{g}(z) = 1 - \left[\frac{\alpha_+}{\gamma_0(2-\beta)}\right]^{\frac{1}{2-\beta}}
\left[\frac{GH_0^2\beta}{\pi\alpha_+}~\Omega_{m0}(1+z)^3\right]^{\frac{1-\beta}{2-\beta}} \,.
\label{fractional DE late time}
\end{align}
This expression is the analytical solution for the dark energy density parameter $\Omega_\mathrm{g}(z)$ for dust matter. 
Applying it at present time, i.e., at $z = 0$, we acquire,
\begin{align}
\Omega_{m0} = \left[\frac{\alpha_+}{\gamma_0(2-\beta)}\right]
\left[\frac{GH_0^2\beta}{\pi\alpha_+}\right]^{1-\beta} \,,
\label{present fractional matter energy}
\end{align}
where we use $1 - \Omega_{g0} = \Omega_{m0}$. 
Using the above expression of $\Omega_{m0}$ into Eq.~(\ref{fractional DE late time}), we obtain,
\begin{align}
\Omega_\mathrm{g}(z) = 1 - \Omega_{m0}\left(1+z\right)^{\frac{3(1-\beta)}{(2-\beta)}} \,.
\label{fractional DE - final}
\end{align}

\begin{figure}[!h]
\begin{center}
\centering
\includegraphics[width=3.5in,height=2.5in]{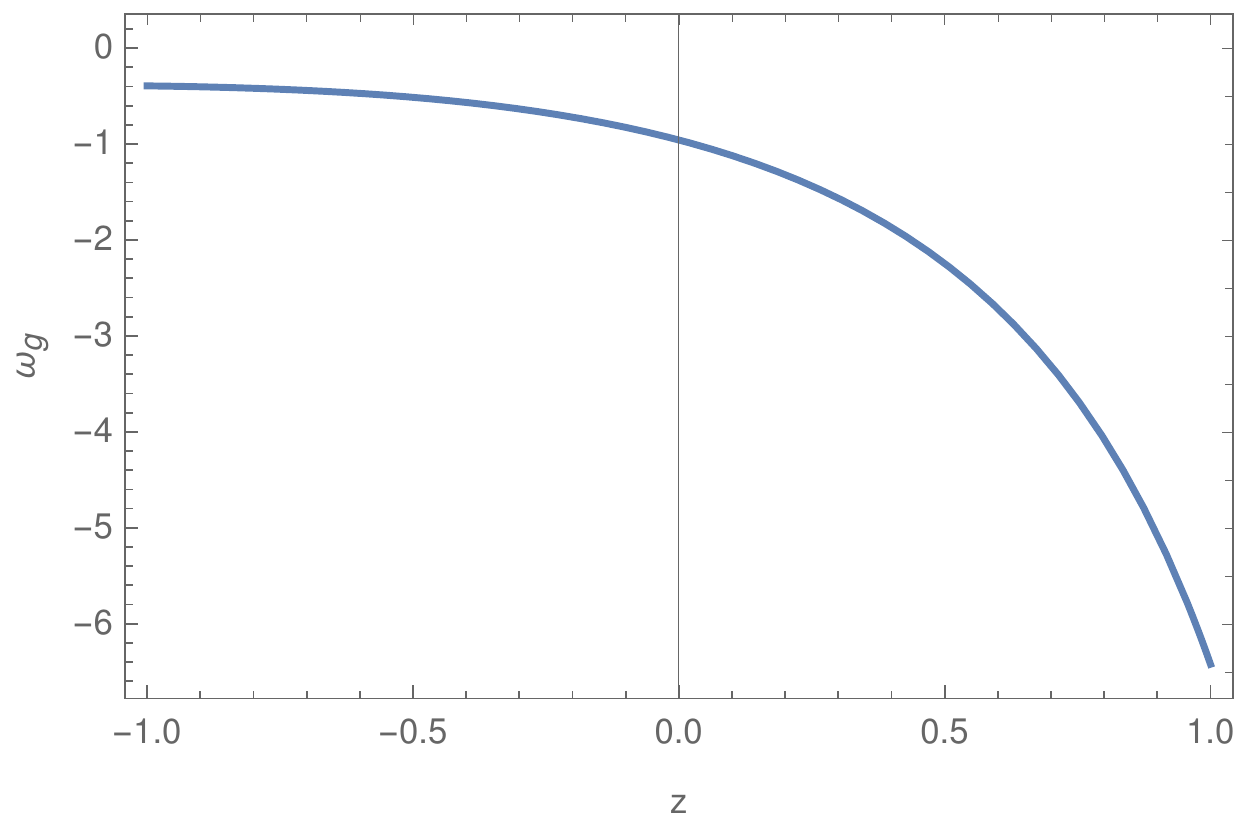}
\caption{$\omega_\mathrm{g}(z)$ vs. $z$ for $\beta = 0.35$, $\left(\alpha_+/\alpha_-\right)^{\beta} = 10$ and $\gamma_0 = 6.5\times10^{-8}$.} 
\label{plot-eos}
\end{center}
\end{figure}

Having all the expressions in hand, we determine the DE EoS parameter from Eq.~(\ref{eos}) and is given by, 
\begin{align}
\omega_\mathrm{g}(z) = -1 + \frac{1}{(2-\beta)}\left(\frac{N}{D}\right) \,,
\label{eos final}
\end{align}
where $N$ (the numerator) and $D$ (the denominator) have the following forms,
\begin{align}
N=&\, 1 - \Omega_{m0}(2-\beta)\left(1+z\right)^{\frac{3(1-\beta)}{(2-\beta)}}
\left\{1 + \left(\frac{\alpha_+}{\alpha_-}\right)^{\beta}\left[\Omega_{m0}(2-\beta)\gamma_0/\alpha_+\right]^{\frac{2\beta}{1-\beta}}
\left(1 + z\right)^{\frac{6\beta}{2-\beta}}\right\} \,,\nonumber\\
D=&\,1 - \Omega_{m0}\left(1+z\right)^{\frac{3(1-\beta)}{(2-\beta)}}
\label{N and D}
\end{align}
respectively. Consequently at $z = 0$,
\begin{align}
\omega_\mathrm{g}(0) = -1 + \frac{1 - \Omega_{m0}(2-\beta)
\left\{1 + \left(\frac{\alpha_+}{\alpha_-}\right)^{\beta}\left[\Omega_{m0}(2-\beta)\gamma_0/\alpha_+\right]^{\frac{2\beta}{1-\beta}}\right\}}
{\left(1 - \Omega_{m0}\right)(2-\beta)} \,.
\label{present eos}
\end{align}
According to the Planck results, the DE EoS parameter at current epoch is constrained by \cite{Planck:2018vyg},
\begin{align}
\omega_\mathrm{g}(0) = -0.957 \pm 0.080 \,.
\label{Planck eos constraint}
\end{align}
For the model at hand, Eq.~(\ref{present eos}) indicates that $\omega_\mathrm{g}(0)$ depends on the parameters 
$\beta$, $\left(\alpha_+/\alpha_-\right)^{\beta}$, $\gamma_0$ and $\alpha_+$. 
Recall that in order to have a $viable$ inflationary scenario during early universe, the parameters should lie within 
the following ranges (as we demonstrated in Eq.~(\ref{inf-constraints})): $\left(\frac{\alpha_+}{\alpha_-}\right)^{\beta} \geq 7.5$, 
$\beta = (0,0.4]$ and $\alpha_+/\beta \approx 10^{-6}$ for $N_f = 58$. 
It may be observed that the parameter $\gamma_0$ remains free from the inflationary constraints, however, it will be constrained from the dark energy requirement. 
In particular, with the aforementioned ranges of $\alpha_+$, $\alpha_-$ and $\beta$, $\omega_\mathrm{g}(0)$ 
becomes compatible with the Planck data provided $\gamma_0$ lies within
\begin{align}
5.5\times10^{-8} \leq \gamma_0 \leq 8\times10^{-8} \,.\label{constraint-gamma}
\end{align}
Considering a particular set of values of the parameters from their viable ranges, 
say $\beta = 0.35$, $\left(\alpha_+/\alpha_-\right)^{\beta} = 10$ and $\gamma_0 = 6.5\times10^{-8}$, we give the plot 
of $\omega_\mathrm{g}(z)$ vs. $z$, see Fig.~\ref{plot-eos}. At this stage it deserves mentioning that the above range of $\gamma_0$ does not support 
the constraint relation between $(\Omega_{m0},H_0)$ in Eq.(\ref{present fractional matter energy}), where we may recall that 
$8\pi GH_0^2 \sim 10^{-120}$. Therefore, in the case of $\Lambda = 0$, the compatibility of $\omega_\mathrm{g}(0)$ with the 
Planck observation and the self-consistency of the constraint Eq.(\ref{present fractional matter energy}) are not concomitantly supported by a
common range of $\gamma_0$.\\

In the next subsection, we will consider the cosmological constant ($\Lambda$) to be non-zero and will examine whether the inclusion of $\Lambda$ 
in the generalized entropic cosmology can lead to a viable dark energy model.

\subsection*{With the cosmological constant: $\Lambda \neq 0$}
In this case, the dark energy density is contributed from the entropic energy density ($\rho_\mathrm{g}$) as well as from the cosmological constant. 
Therefore the dark energy density and the corresponding pressure can be written as,
\begin{eqnarray}
 \rho_\mathrm{D}&=&\rho_\mathrm{g} + \frac{3}{8\pi G}\left(\frac{\Lambda}{3}\right)~~,\nonumber\\
 \rho_\mathrm{D} + p_\mathrm{D}&=&\rho_\mathrm{g} + p_\mathrm{g}~~,
 \label{de-1}
\end{eqnarray}
with $\rho_\mathrm{g}$ and $p_\mathrm{g}$ are given in Eq.(\ref{energy and pressure-late time}), and recall that 
the cosmological constant has the equation of state like $\rho_\mathrm{\Lambda} + p_\mathrm{\Lambda} = 0$. 
Consequently, the dark energy EoS parameter comes with the following expression:
\begin{align}
\omega_\mathrm{D} = p_\mathrm{D}/\rho_\mathrm{D} = -1 - \left(\frac{2\dot{H}}{3H^2}\right)
\left[\frac{1 - \frac{\alpha_+}{\gamma_0}\left(\frac{GH^2\beta}{\pi\alpha_+}\right)^{1-\beta} 
 - \left(\frac{\alpha_+}{\gamma_0}\right)\left(\frac{\alpha_+}{\alpha_-}\right)^{\beta}\left(\frac{GH^2\beta}{\pi\alpha_+}\right)^{1+\beta}}
{1 - \frac{\alpha_+}{\gamma_0(2-\beta)}\left(\frac{GH^2\beta}{\pi\alpha_+}\right)^{1-\beta} + \frac{\Lambda}{3H^2}}\right] \,.
\label{eos-1}
\end{align}
In presence of the cosmological constant, the Friedmann equations are written as,
\begin{align}
H^2 = \frac{8\pi G}{3}\left(\rho_m + \rho_\mathrm{D}\right) 
= \frac{8\pi G}{3}\left(\rho_m + \rho_\mathrm{g}\right) + \frac{\Lambda}{3} \,,\nonumber\\
\dot{H} = -4\pi G\left[\rho_m + \left(\rho_\mathrm{D} + p_\mathrm{D}\right)\right] 
= -4\pi G\left[\rho_m + \left(\rho_\mathrm{g} + p_\mathrm{g}\right)\right] \,.
\label{FRW-late time-1}
\end{align} 
As usual, the fractional energy density of the pressureless matter and the dark energy satisfy $\Omega_m + \Omega_\mathrm{D} = 1$ which along with 
$\rho_m = \rho_{m0}\left(\frac{a_0}{a}\right)^3$ (with $\rho_{m0}$ being 
the present matter energy density) result to the Hubble parameter in terms of the red shift factor ($z$) as follows,
\begin{align}
H(z) = \frac{H_0\sqrt{\Omega_{m0}(1+z)^3}}{\sqrt{1-\Omega_\mathrm{D}}} \,.
\label{Hubble late time-1}
\end{align}
The presence of a non-vanishing cosmological constant modifies the expression of $H(z)$ through the factor $\Omega_\mathrm{D}$ compared to the previous 
case when $\Lambda = 0$. Plugging the expression of $\rho_\mathrm{g}$ from Eq.~(\ref{energy and pressure-late time}) into 
$\Omega_\mathrm{D} = \left(\frac{8\pi G}{3H^2}\right)\rho_\mathrm{g} + \frac{\Lambda}{3}$, and using the above form of $H(z)$, we obtain,
\begin{align}
\Omega_\mathrm{D}(z) = 1 - \frac{\left[\frac{\alpha_+}{\gamma_0(2-\beta)}\right]^{\frac{1}{2-\beta}}
\left[\frac{GH_0^2\beta}{\pi\alpha_+}~\Omega_{m0}(1+z)^3\right]^{\frac{1-\beta}{2-\beta}}}
{\left[1 + \frac{\Lambda}{3H_0^2\Omega_{m0}(1+z)^3}\right]^{1/(2-\beta)}}\,.
\label{fractional DE late time-1}
\end{align}
Clearly for $\Lambda = 0$, the above expression of $\Omega_\mathrm{D}(z)$ resembles with that of the earlier case (see Eq.(\ref{fractional DE late time})). 
At $z = 0$, Eq.(\ref{fractional DE late time-1}) leads to,
\begin{align}
\Omega_{m0}  + \frac{\Lambda}{3H_0^2} = \left[\frac{\alpha_+}{\gamma_0(2-\beta)}\right]
\left[\frac{GH_0^2\beta}{\pi\alpha_+}\right]^{1-\beta} \,,
\label{present fractional matter energy-1}
\end{align}
which provides a constraint relation between $(\Lambda,\Omega_{m0},H_0)$, i.e Eq.(\ref{present fractional matter energy-1}) relayes 
the $\Lambda$ with  the observationally determined quantities $\Omega_{m0}$ and $H_0$. 
Using the above expression of $\Omega_{m0}$ into Eq.~(\ref{fractional DE late time-1}), we obtain,
\begin{align}
\Omega_\mathrm{D}(z) = 1 - \Omega_{m0}\left(1+z\right)^{\frac{3(1-\beta)}{(2-\beta)}}f(\Lambda,\Omega_{m0},H_0,z) \,,
\label{fractional DE - final-1}
\end{align}
where the function $f(\Lambda,\Omega_{m0},H_0,z)$ has the following form:
\begin{eqnarray}
 f(\Lambda,\Omega_{m0},H_0,z) = \left\{\frac{1 + \frac{\Lambda}{3H_0^2\Omega_{m0}}}{1 + \frac{\Lambda}{3H_0^2\Omega_{m0}(1+z)^3}}\right\}^{1/(2-\beta)}~~.
 \label{f}
\end{eqnarray}
The function $f(\Lambda,\Omega_{m0},H_0,z)$ actually encodes the information of $\Lambda$, and thus the presence of the cosmological constant 
non-trivially affects the dark energy density parameter by Eq.(\ref{fractional DE - final-1}). Therefore it is important to examine that 
whether the inclusion of the cosmological constant in the generalized entropic cosmology can resolve the deficiency faced in the case of 
$\Lambda = 0$, and leads to a viable dark energy model of the universe.  

\begin{figure}[!h]
\begin{center}
\centering
\includegraphics[width=3.5in,height=2.5in]{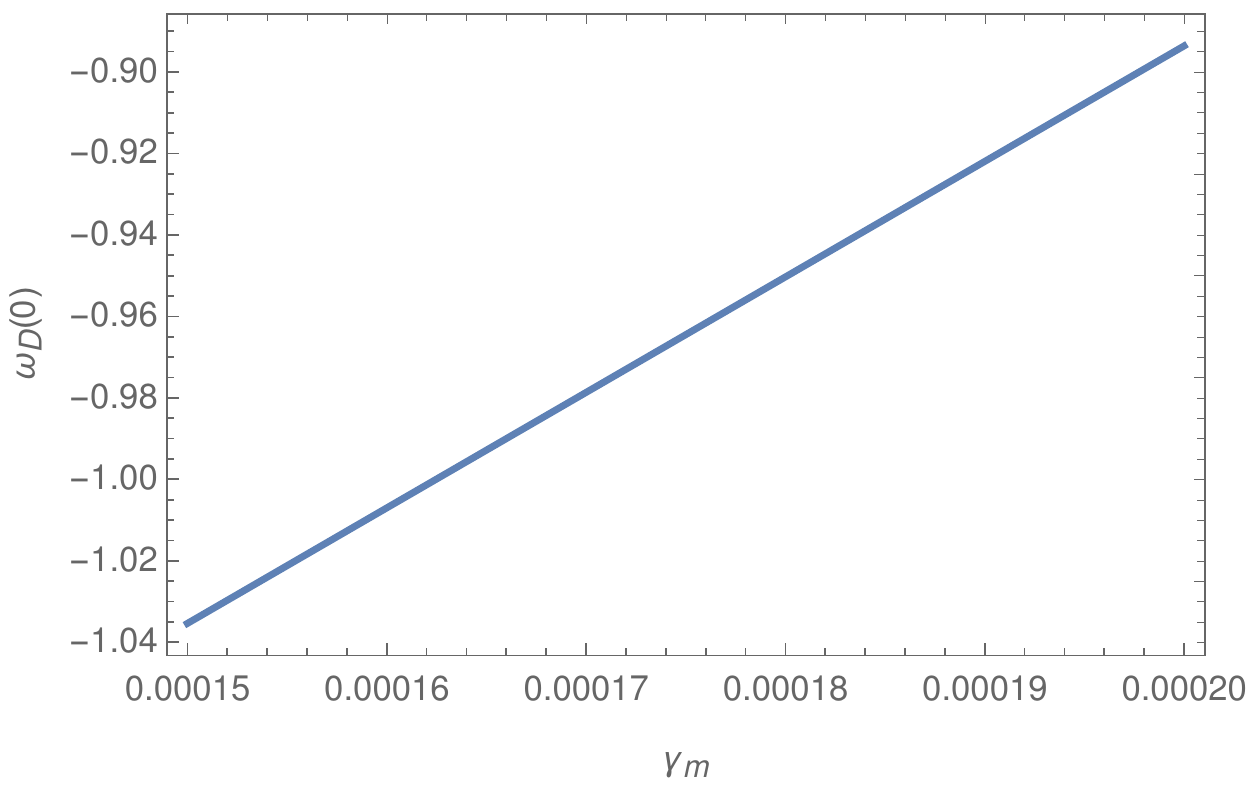}
\caption{$\omega_\mathrm{D}(0)$ vs. $\gamma_m$ 
for $\beta = 0.35$, $\left(\alpha_+/\alpha_-\right)^{\beta} = 10$, $\alpha_{+}/\beta = 10^{-6}$ (according to Eq.(\ref{inf-constraints})) 
and $8\pi GH_0^2 = 10^{-120}$.} 
\label{plot-new1}
\end{center}
\end{figure}

By using the above expressions, we determine the DE EoS parameter from Eq.~(\ref{eos-1}) as follows, 
\begin{align}
\omega_\mathrm{D}(z) = -1 + \frac{1}{(2-\beta)\left(1 + \frac{\Lambda}{3H_0^2\Omega_{m0}(1+z)^3}\right)}\left(\frac{N}{D}\right) \,,
\label{eos final-1}
\end{align}
where $N$ (the numerator) and $D$ (the denominator) have the following forms,
\begin{eqnarray}
N=1 - \Omega_{m0}(2-\beta)\left(1+z\right)^{\frac{3(1-\beta)}{(2-\beta)}}
\Bigg\{\left(\frac{1+\frac{\Lambda}{3H_0^2\Omega_{m0}}}{\left[f(\Lambda,\Omega_{m0},H_0,z)\right]^{1-\beta}}\right) 
&+&\left(\frac{\alpha_+}{\alpha_-}\right)^{\beta}\left[\Omega_{m0}(2-\beta)\gamma_0/\alpha_+\right]^{\frac{2\beta}{1-\beta}}
\left(1 + z\right)^{\frac{6\beta}{2-\beta}}\nonumber\\
&\times&\left(\frac{\left[1+\frac{\Lambda}{3H_0^2\Omega_{m0}}\right]^{(1+\beta)/(1-\beta)}}
{\left[f(\Lambda,\Omega_{m0},H_0,z)\right]^{1+\beta}}\right)\Bigg\} \,,
\nonumber
\end{eqnarray}
and
\begin{eqnarray}
D = 1 - \Omega_{m0}\left(1+z\right)^{\frac{3(1-\beta)}{(2-\beta)}}\left(\frac{1+\frac{\Lambda}{3H_0^2\Omega_{m0}}}
{\left[f(\Lambda,\Omega_{m0},H_0,z)\right]^{1-\beta}}\right) + \frac{\Lambda}{3H_0^2}\left(\frac{f(\Lambda,\Omega_{m0},H_0,z)}
{(1+z)^{3/(2-\beta)}}\right)
\label{N and D-1}
\end{eqnarray}
respectively. Consequently at $z = 0$,
\begin{align}
\omega_\mathrm{D}(0) = -1 + \frac{1 - \Omega_{m0}(2-\beta)
\left\{\left(1+\frac{\Lambda}{3H_0^2\Omega_{m0}}\right) 
+ \left(\frac{\alpha_+}{\alpha_-}\right)^{\beta}\left[\Omega_{m0}(2-\beta)\gamma_0/\alpha_+\right]^{\frac{2\beta}{1-\beta}}
\left(1+\frac{\Lambda}{3H_0^2\Omega_{m0}}\right)^{(1+\beta)/(1-\beta)}\right\}}
{\left(1 - \Omega_{m0}\right)(2-\beta)\left(1+\frac{\Lambda}{3H_0^2\Omega_{m0}}\right)} \,.
\label{present eos-1}
\end{align}
Recall that $\Lambda$, $\Omega_{m0}$ and $H_0$ are related by Eq.(\ref{present fractional matter energy-1}), due to which, we may write $\Lambda$ 
in terms of the model parameters and $(\Omega_{m0},H_0)$,
\begin{eqnarray}
1+\frac{\Lambda}{3H_0^2\Omega_{m0}} = \frac{\alpha_{+}}{(2-\beta)\Omega_{m0}}\left(\frac{\beta}{8\pi^2\alpha_{+}}\right)^{1-\beta}
\left[\frac{\left(8\pi GH_0^2\right)^{1-\beta}}{\gamma_0}\right]~~.
\label{relation-1}
\end{eqnarray}
Therefore $\omega_\mathrm{D}(0)$ depends on the parameters: 
$\beta$, $\left(\alpha_+/\alpha_-\right)^{\beta}$, $\gamma_0$ and $\alpha_+$. The inflationary quantities like the scalar spectral index and the tensor 
to scalar ratio in the present context are found to be simultaneously compatible with the Planck data if some of the parameters 
like $\alpha_{+}$, $\alpha_{-}$ and $\beta$ get constrained according to Eq.~(\ref{inf-constraints}), while the parameter $\gamma_0$ remains 
free from the inflationary requirement. With the aforementioned ranges of $\alpha_+$, $\alpha_-$ and $\beta$, $\omega_\mathrm{D}(0)$ 
becomes compatible with the Planck observational data (see Eq.(\ref{Planck eos constraint})), 
provided $\gamma_0$ lies within a small window as follows,
\begin{align}
1.5\times10^{-4} \leq \frac{\gamma_0}{\left(8\pi GH_0^2\right)^{1-\beta}} \leq 2\times10^{-4} \,,\label{constraint-gamma-1}
\end{align}
which, for a certain value of $\beta$ and with $8\pi GH_0^2 \sim 10^{-120}$, provides a viable constraint on $\gamma_0$. This is depicted in 
Fig.~\ref{plot-new1} which clearly demonstrates that the theoretical expectation of $\omega_\mathrm{D}(0)$ gets consistent with the Planck 
observational data for $\gamma_m = [1.5\times10^{-4},2\times10^{-4}]$ where $\gamma_m = \gamma_0/\left(8\pi GH_0^2\right)^{1-\beta}$.

\begin{figure}[!h]
\begin{center}
\centering
\includegraphics[width=3.5in,height=2.5in]{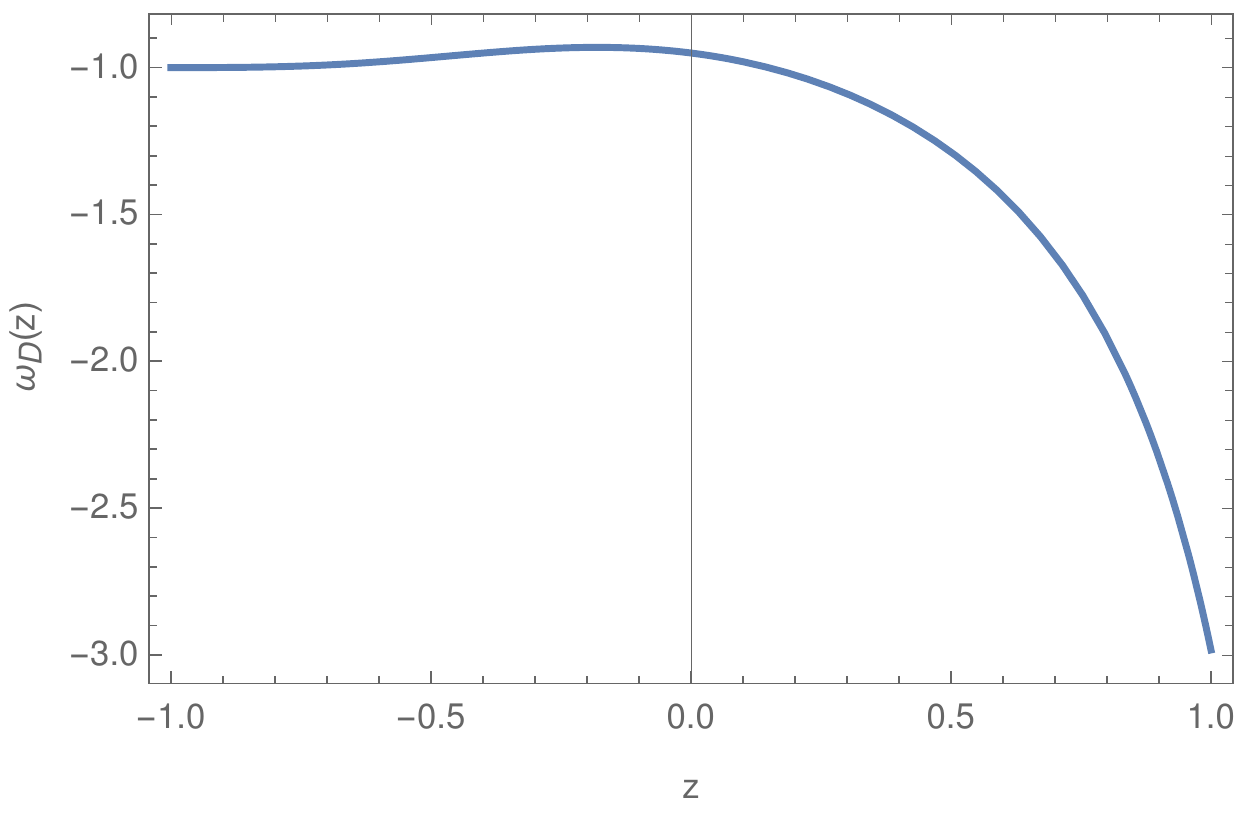}
\caption{$\omega_\mathrm{D}(z)$ vs. $z$ for a particular set of values of the parameters from their viable ranges 
as per Eq.(\ref{inf-constraints}) and Eq.(\ref{constraint-gamma-1}), say 
$\beta = 0.35$, $\left(\alpha_+/\alpha_-\right)^{\beta} = 10$, $\alpha_{+}/\beta = 10^{-6}$ and $\gamma_m = 1.8\times10^{-4}$.} 
\label{plot-eos1}
\end{center}
\end{figure}

Furthermore the deceleration parameter (symbolized by $q$) at present universe is obtained as,
\begin{eqnarray}
 q = -1 + \frac{3}{2(2-\beta)\left(1 + \frac{\Lambda}{3H_0^2\Omega_{m0}}\right)}~~.
 \label{q}
\end{eqnarray}
Therefore for $\gamma_m = [1.5\times10^{-4},2\times10^{-4}]$, the theoretical expression of $q$ lies within 
$q = [-0.56,-0.42]$ which certainly contains the observational value of $q = -0.535$ from the Planck data \cite{Planck:2018vyg}. In particular, 
$q = -0.535$ occurs for $\gamma_m = 1.8\times10^{-4}$. Considering this value of $\gamma_m$ and by using 
Eq.(\ref{eos final-1}), we give the plot of $\omega_\mathrm{D}(z)$ vs. $z$, see Fig.~\ref{plot-eos1}. 
The figure reveals that that the theoretical expectation of the 
DE EoS parameter at present time acquires the value: $\omega_\mathrm{D}(0) = -0.950$ which is well consistent with the Planck observational 
data \cite{Planck:2018vyg}.
 
Therefore in the case of $\Lambda \neq 0$ where the dark energy density consists of the generalized entropic energy density and the cosmological constant, 
we have found that the theoretical expectation of $\omega_\mathrm{D}(0)$ gets compatible with the Planck data for suitable ranges of the 
model parameters, and for such ranges, the constraint relation between $(\Lambda,\Omega_{m0},H_0)$ becomes self-consistent. This is unlike to the 
case of $\Lambda = 0$ where the compatibility of $\omega_\mathrm{D}(0)$ with the Planck observation does not support the 
constraint relation between $(\Omega_{m0},H_0)$. Moreover, by comparing Fig.~\ref{plot-eos} and Fig.~\ref{plot-eos1}, we may argue that 
the dark energy EoS parameter in presence of the cosmological constant acquires less negative values around $z = 1$ compared to the case when the 
cosmological constant is absent. Therefore the inclusion of an explicit cosmological constant improves the cosmological behaviour of the generalized 
entropic cosmology in the dark energy sector, and leads to a more viable entropic dark energy model compared to the case of $\Lambda = 0$. 
Such effects of cosmological constant in the context of entropic cosmology is in agreement with \cite{Lymperis:2018iuz} where the authors constructed a modified 
cosmological scenario from non-extensive Tsallis entropy function with / without a cosmological constant.\\ 

As a whole, we may argue that the entropic cosmology from the generalized entropy function $S_\mathrm{g}$ can 
unify the early inflation to the late 
dark energy era of the universe, for suitable ranges of the parameters given by:
\begin{align}
\sigma_0=&\,[0.013,0.017] \,, \quad \left(\alpha_+/\alpha_-\right)^{\beta} \geq 7.5 \,,\nonumber\\
\beta=&\, (0,0.4]\ \mbox{and}\ \gamma_m = [1.5\times10^{-4},2\times10^{-4}] \,.
\label{constraints-final}
\end{align}
In such a unified scenario, we find that -- (1) the inflation is described by a quasi de-Sitter evolution of the 
Hubble parameter, which has an exit mechanism at around 58 e-folding number, (2) the inflationary observable quantities like the spectral index 
for primordial scalar perturbation and the tensor-to-scalar ratio are simultaneously compatible with the recent Planck data, and 
(3) regarding the late time cosmology, the dark energy EoS parameter gets consistent with the Planck results for the same values of the parameters 
that lead to the viable inflation during the early universe. 
Moreover, it may be noticed that for the parameter ranges shown in Eq.(\ref{constraints-final}), 
the form of the $S_\mathrm{g}$ does not tend to any of the known entropies. Therefore 
the present work clearly depicts that the unification of early inflation to the dark energy in holographic cosmology 
demands an entropy function other than the 
Tsallis, R\'{e}nyi, Barrow, Sharma-Mittal and Kaniadakis entropies. Thus it is very important to find the true entropy function that provides a 
viable unification of inflation with the dark energy era of the universe in the context of holographic
cosmology, and we have carried this investigation in the current work.\\

Before concluding, here it may be mentioned that it will be an interesting avenue to apply the generalized entropic function in describing 
an alternative early universe scenario, in particular the bouncing scenario. 
Recently a unified cosmological model has been proposed from a non-singular ekpyrotic 
bounce to the dark energy era with an intermediate matter-dominated epoch, and the primordial as well 
as the dark energy observable quantities are found to be compatible with the latest Planck constraints \cite{Nojiri:2022xdo}. 
The possibility of such a unified scenario from the entropic function proposed in the present work will be examined elsewhere.

\section{Conclusion}

We propose a new four-parameter entropy function ($S_\mathrm{g}$) that generalizes the Tsallis, R\'{e}nyi, Barrow, Sharma-Mittal, 
Kaniadakis and Loop Quantum Gravity entropies for suitable limits of the parameters. 
The generalized entropy function satisfies the third law of thermodynamics, i.e., $S_\mathrm{g}$ goes to zero 
at $S \rightarrow 0$ (where $S$ represents the Bekenstein-Hawking entropy), and moreover 
$S_\mathrm{g}$ seems to reduce to the Bekenstein-Hawking entropy at a certain limit. 
In this regard, we give the following postulate -- 
``The minimum number of parameters required in a generalized entropy function that can generalize all the aforementioned entropies is equal to four''. 
The effective energy density and the effective pressure sourced from the $S_\mathrm{g}$ modify the 
Friedmann equations, and consequently, we address the early and late time cosmology of the universe corresponding to the 
generalized four-parameter entropy function. During the early universe, the cosmological constant is highly suppressed in respect to the entropic energy 
density and thus we can safely ignore the cosmological constant in studying the early stage of the universe. Therefore the effective energy density 
during early universe solely arises from the entropic energy density which in turn triggers a successful inflationary era with an exit. 
In particular, the early phase of the universe is described by a quasi de-Sitter inflationary era, 
and the typical energy scale of the universe at the onset of inflation becomes of the order $\sim 10^{-3}M_\mathrm{Pl}$. 
We calculate the slow roll parameter during the inflation (in terms of the e-folding number symbolized by $N$), which remains less than unity 
and almost constant for $N<N_f = 58$ and rises to unity at $N=N_f$. 
This indicates that the universe gets an exit from the inflation at around 58 e-folding number. 
To examine the viability of the inflationary scenario, we determine the primordial observable quantities 
like the spectral index for curvature perturbation and the tensor-to-scalar ratio. 
As a result, the scalar spectral index and the tensor-to-scalar ratio turn out to be 
simultaneously compatible with the recent Planck data for suitable ranges of the parameters present in the entropy function ($S_\mathrm{g}$). 
Regarding the late time cosmology, we find that the energy density of the generalized entropy triggers a late time accelerating phase of the universe, 
and thus acts as a candidate of the dark energy density. 
Here we would like to mention that although the cosmological constant does not play any role in the early-time universe, it contributes a significant role 
during the late dark energy epoch of the universe. Thus in regard to the dark energy era, we investigate two distinct cases, 
namely when the explicit cosmological constant ($\Lambda$) is absent and when it is present. For $\Lambda = 0$, the dark energy density is solely 
contributed from the entropic energy density, while for $\Lambda \neq 0$, the dark energy density consists of the entropic energy density and the 
cosmological constant as well. In both the cases, we analytically determine the dark energy EoS parameter of the universe, and they 
are found to be compatible with the Planck data for that values of the parameters which lead to the viable inflation during the early universe. However 
in presence of the cosmological constant, the entropic dark energy model turns out to be 
more viable compared to the case of $\Lambda = 0$. Actually in presence 
of the cosmological constant, the compatibility of the dark energy EoS parameter with the Planck observation leads to the self-consistency 
of the constraint relation between $(\Lambda,\Omega_{m0},H_0)$, however for $\Lambda = 0$, the compatibility of $\omega_\mathrm{D}(0)$ with the Planck data 
does not support the constraint relation between $(\Omega_{m0},H_0)$. 
Moreover in the case of $\Lambda \neq 0$, the deceleration parameter ($q$) at present universe also seems to support the Planck data. 

As a whole, in the present context of 
entropic cosmology where the generalized entropy function contains four independent parameters, 
the inflationary quantities like $(n_s,r)$ and the dark energy quantities like $(\omega_\mathrm{D},q)$ are found to be 
simultaneously satisfied for suitable range of the parameters.
Therefore we may argue that the entropic cosmology from the generalized entropy function ($S_\mathrm{g}$) 
is able to unify the early inflation to the late dark energy era of the universe. 
Furthermore, we show that the entropic cosmology from the proposed 
entropy is equivalent to holographic cosmology with suitable forms of holographic cut-offs. In particular, the holographic cut-offs 
are determined in terms of either particle horizon and its derivative or future horizon and its derivative. 

\section*{Acknowledgments}

This work was supported in part by MINECO (Spain), project PID2019-104397GB-I00 (SDO). 
This work was partially supported by the program Unidad de Excelencia Maria 
de Maeztu CEX2020-001058-M. This research was also supported in part by the 
International Centre for Theoretical Sciences (ICTS) for the online program - Physics of the Early Universe (code: ICTS/peu2022/1) (TP).

\end{document}